\documentclass[preprint,showpacs,preprintnumbers,superscriptaddress]{revtex4-1}
\usepackage{multirow}
\usepackage{diagbox}
\usepackage{booktabs}
\usepackage{array}
\usepackage{natbib}
\usepackage{amsmath,amssymb,graphicx,bm,subfigure,float,siunitx,color}
\usepackage{bm,subfigure,float,siunitx,graphicx}
\usepackage[figuresright]{rotating}
\usepackage{color}
\usepackage{epstopdf}
\usepackage{mathrsfs}
\newcommand{\be}{\begin{equation}}
\newcommand{\ee}{\end{equation}}

\newcommand{\1}{\left}
\newcommand{\2}{\right}
\def\({\left(}
\def\){\right)}
\def\[{\left[}
\def\]{\right]}

\newcommand{\dif}{\,\mathrm{d}}

\newcommand{\me}{\mathrm{e}}

\newcommand{\p}{\partial}

\newcommand{\m}{\mu}
\newcommand{\n}{\nu}
\newcommand{\al}{\alpha}
\newcommand{\bet}{\beta}

\newcommand{\lam}{\lambda}
\newcommand{\sig}{\sigma}

\newcommand{\om}{\omega}

\renewcommand{\th}{\theta}

\newcommand{\na}{\nabla}

\begin{document}
\title{\boldmath Choked accretion onto Kerr-Sen black holes in Einstein-Maxwell-dilaton-axion gravity}
\author{Haiyuan Feng \footnote{Corresponding author}}
\email{Email address: 406606114@qq.com}
\affiliation{Department of Physics, Southern University of Science and Technology, Shenzhen 518055, Guangdong, China}

\author{Yingdong Wu}
\email{Email address: 12131274@mail.sustech.edu.cn}
\affiliation{Department of Physics, Southern University of Science and Technology, Shenzhen 518055, Guangdong, China}

\author{Rong-Jia Yang \footnote{Corresponding author}}
\email{Email address: yangrongjia@tsinghua.org.cn}
\affiliation{College of Physical Science and Technology, Hebei University, Baoding 071002, China}

\author{Leonardo Modesto}
\email{Email address: lmodesto@sustech.edu.cn}
\affiliation{Department of Physics, Southern University of Science and Technology, Shenzhen 518055, Guangdong, China}

\begin{abstract}
We investigate the process of an ultrarelativistic fluid accreted onto axisymmetric Kerr-Sen black holes in Einstein-Maxwell-dilaton-axion theory. We obtain the solution describing the velocity potential of a stationary irrotational fluid with an stiff equation of state and the solution for the streamlined diagram of the quadrupolar flow. We also investigate how the solution's coefficients and the stagnation points are affected by the parameters. The injection rate, the ejection rate, and the critical angle are discussed in detail. We find that with an increasing dilation parameter the ratio of the ejection rate to the injection rate increases and that the radiative efficiency is larger, while the redshift is lower, compared to the Kerr black hole.
 \end{abstract}

\maketitle

\section{Introduction}
Accretion of matter onto a black hole is the most probable scenario for explaining the high energy output from active galactic nuclei and quasars. It is also provides the most credible expositions for the high energy outflow from X-ray Binaries and Gamma Ray Bursts \cite{Abramowicz:2011xu}. The initial exploration of accretion process onto celestial objects were investigated by Hoyle and Lyttleton \cite{hoyle_lyttleton_1939,hoyle_lyttleton_1940_1,hoyle_lyttleton_1940_2,1941MNRAS.101..227H}. Subsequently, Bondi and Hoyle focused on the accretion of gas without pressure onto a moving star \cite{Bondi:1944jm}. Bondi formulated the theory of the transonic hydrodynamic accretion of the adiabatic fluid in a stationary, spherically symmetric spacetime \cite{Bondi:1952ni}. This scheme for accretion process was generalized to the Schwarzschild black hole by Michel \cite{Michel1972}. Petrich formulated a relativistic framework for the stationary accreted fluid with an adiabatic equation of state onto a Schwarzschild and Kerr black hole \cite{Petrich:1988zz}. In recent developments, researchers considered the accretion of Vlasov gas, ignoring the back reaction of matter, and analyzed its accretion rate \cite{Mach:2021zqe, Mach:2020wtm, Cai:2022fdu}. Since then, a considerable amount of literature have been dedicated to theoretical and observational studies for accretion. \cite{Malec:1999dd, Babichev:2004yx, Jiao:2016iwp, Ganguly:2014cqa, Yang:2015sfa, Font:1998sc, Yang:2019qru, Parev:1995me, Font:1998sc,DiMatteo:2002hif, Silich:2008xz, Gil-Merino:2011yhj, Kuo:2014pqa, He:2022lrc}.%
On the flip side, the astrophysical jet represents another phenomenon, where plasma is emitted in the form of an extended beam along the rotation axis.
The jet mechanism proposed by Blandford and Payne (BP) and Blandford and Znajek (BZ) stand as the most widely accepted explanations in prior studies \cite{10.1093/mnras/199.4.883, 10.1093/mnras/179.3.433}. BP mechanism utilizes magnetic centrifugation to extract energy and angular momentum from a black hole accretion disc, whereas the BZ mechanism involves rotational energy from a Kerr black hole. These methods were commonly used in magneto-hydrodynamics and astrophysics explorations \cite{Semenov:2004ib,McKinney:2006tf,Qian:2018qpg,Liska:2018ayk}. Afterward, a new hydrodynamic accretion mechanism was proposed to explain the effect of the steady-state axisymmetric partial fluid flowing from the equatorial plane being ejected along poles by the cytaster's gravitational influence \cite{https://doi.org/10.48550/arxiv.1103.0250}. This process was also generalized to the choked accretion scenario of Schwarzschild and Kerr black holes \cite{Tejeda:2019fwr,Aguayo-Ortiz:2020qro}.

The preceding discussions were primarily confined to the general relativity (GR) framework. Similar accretion and jet phenomena exist for black holes in alternative gravity models \cite{Jiao:2016uiv,Yang:2021opo,Medvedev:2002wn,Fragile:2004gp,Bhadra:2011me,JimenezMadrid:2005rk,Feng:2022bst}. In GR, the theoretical predictability is broken by singularities associated with black holes and by the big bang \cite{PhysRevLett.14.57,PhysRevD.14.2460,Christodoulou:1991yfa}. This suggests that GR may not be a complete theory of gravity, highlighting the necessity for modifications from a more comprehensive theory that incorporates the quantum nature at very small length scales \cite{Will:2014kxa}. From an observational standpoint, GR encounters challenges in understanding the characteristics of dark matter and dark energy, both essential for explaining phenomena such as the rotation curves of galaxies and the accelerated expansion of the universe, respectively \cite{1983ApJ...270..371M,1984ApJ...286....7B,Milgrom:2003ui,SupernovaSearchTeam:1998fmf,Clifton:2011jh}. Furthermore, recent developments suggest a breakdown of GR in low-acceleration regime, thereby exerting profound influence on both the astrophysical and cosmological domains \cite{hernandez2023statistical,chae2023robust}.

Among many alternative theories to traditional GR, the Einstein-Maxwell-dilaton-axion (EMDA) model has attracted significant attentions \cite{Rogatko:2002qe,Sen:1992ua}. The model incorporates the dilation field and the pseudoscalar axion, both of which are connected to the metric and the Maxwell field. The origins of the dilaton and axion fields can be attributed to string compactifications, giving rise to compelling implications in the inflationary and late-time accelerated cosmologies \cite{Catena:2007jf,Sonner:2006yn}. Therefore, it is valuable to investigate the role of such a theory in astrophysical observations. Within these string-inspired low-energy effective theories, the parameters were constrained from observations, for instance, a preferred value of $r_2\equiv\frac{Q^2}{M}\thickapprox0.2M$ is determined based on the optical continuum spectrum of quasars \cite{Banerjee:2020qmi}. Additionally, a recent investigation gave an observational constraint on the dilaton parameter ($0.1M\lesssim r_2\lesssim0.4M$) by analyzing the shadow diameters of M87* and Sgr A* \cite{Sahoo:2023czj}. Finally, a constraint on the dilaton parameter is obtained by employing simulated data that replicates potential observations of the S2 star via a gravity interferometer \cite{Fernandez:2023kro}. It demonstrated that enhanced astrometric accuracy can effectively narrow down the acceptable range of dilaton parameter to $r_2 \lesssim0.066M$.

The article is organized as follows: In Section II, the EMDA model and the Kerr-Sen black hole will be briefly reviewed. In Section III, for irrotational fluids, we will derive an analytical solution for the velocity potential $\Phi$ in the Boyer-Lindquist coordinate system. In Section IV, the quadrupolar flow solution will be used to analyze the variation of the coefficients with parameters, and the streamline and the temperature diagrams will be displayed in the ZAMO frame. In Section V, we will present the mechanism of choked accretion, including the density ratio, the injection rate, the ejection rate, as well as the value range of the reference point. Finally, we will analyze the variation of radiative efficiency $\epsilon$ and redshift $z$ with different parameters. For convenience, we will use geometrical units $c=G=1$ and the signature convention $(-,+,+,+)$ for the spacetime metric throughout the article.

\section{Kerr-Sen black hole in Einstein-Maxwell-dilaton-axion gravity}
The EMDA model is derived from the low energy limit behavior of heterotic string theory. it is composed of dilaton field $\chi$, gauge vector field $A_{\m}$, metric $g_{\m\n}$, and pseudo-scalar axion field $\xi$ \cite{Sen:1992ua,Rogatko:2002qe,Campbell:1992hc}. The action of the EMDA model can be formulated through the coupling of supergravity and super-Yang Mills theory, and it can be described by the following form
\be
\label{1}
S=\frac{1}{16\pi}\int\sqrt{-g}d^4x\[\tilde{R}-2\p_{\m}\chi\p^{\m}\chi-\frac{1}{2}\me^{4\chi}\p_{\m}\xi\p^\m\xi+
\me^{-2\chi}F_{\m\n}F^{\m\n}+\xi F_{\m\n}\widetilde{F}^{\m\n} \],
\ee
where $\tilde{R}$ is the Ricci scalar, $F_{\m\n}$ is the second-order antisymmetric Maxwell field strength tensor with $F_{\m\n}=\nabla_{\m}A_{\n}-\nabla_{\n}A_{\m}$, and $\widetilde{F}^{\m\n}$ is the dual tensor of the field strength. The variation of the aforementioned four fields yields the following motion equations:
\be
\1\{\begin{split}
\label{2}
&\Box\chi-\frac{1}{2}\me^{4\chi}\na_\m\xi\na^\m\xi+\frac{1}{2}\me^{-2\chi}F_{\m\n}F^{\m\n}=0,\\
&\Box\xi+4\na_\m\xi\na^\m\xi-\me^{-4\chi}F_{\m\n}\widetilde{F}^{\m\n}=0,\\
&\na_\m\widetilde{F}^{\m\n}=0,\\
&\na_\m(\me^{-2\chi}F^{\m\n}+\xi\widetilde{F}^{\m\n})=0,\\
&G_{\m\n}=\me^{2\chi}(4F_{\m\rho}F^{\rho}_{\n}-g_{\m\n}F^2)-g_{\m\n}(2\na_\m\chi\na^\m\chi+\frac{1}{2}\me^{4\chi}\na_\m\xi\na^\m\xi),\\
&+\na_\m\chi\na_\n\chi+\me^{4\chi}\na_\m{\xi}\na_\n{\xi}.
\end{split}\2.
\ee
These equations indicate that the dilaton field, the axion field, the electromagnetic field, and the gravitational field are observed to be coupled. The classical axisymmetric solution, known as the Kerr-Sen solution, can be expressed in Boyer-Lindquist coordinates as \cite{Garcia:1995qz, Ghezelbash:2012qn, Bernard:2016wqo}.
\be
\label{3}
\begin{split}
\dif s^{2}&=-\(1-\frac{2M r}{\tilde{\Sigma}}\)\dif t^{2}+\frac{\tilde{\Sigma}}{\tilde{\Delta}}\dif r^{2}+\tilde{\Sigma}\dif \th^{2}-\frac{4aMr}{\tilde{\Sigma}}\sin^2{\th}\dif t\dif\phi\\
&+\sin^2{\th}\dif \phi^2\(r(r+r_2)+a^2+\frac{2Mra^2\sin^2{\th}}{\tilde{\Sigma}}\),
\end{split}
\ee
with
\be
\1\{\begin{split}
\label{4}
&\tilde{\Sigma}=r(r+r_2)+a^2\cos^2{\th},\\
&\tilde{\Delta}=r(r+r_2)-2Mr+a^2,
\end{split}\2.
\ee
where $M$ is the mass parameter of the black hole, the dilaton parameter is defined as $r_2=\frac{Q^2}{M}$ ($Q$ represents the electric charge) and $a$ denotes the black hole's angular momentum per unit mass. Eq. \eqref{3} indicates that when the black hole's rotation parameter is excluded, it results in a spherically symmetric dilaton black hole composed of mass, electric charge, and asymptotic value \cite{Garfinkle:1990qj}. When the dilaton parameter $r_2$ vanishes, the Kerr-Sen solution reverts to the Kerr black hole.

The event horizon $r_\pm$ of the Kerr-Sen black hole is determined by
\be
\1\{\begin{split}
\label{5}
&r_+=M-\frac{r_2}{2}+\sqrt{\(M-\frac{r_2}{2}\)^2-a^2},\\
&r_-=M-\frac{r_2}{2}-\sqrt{\(M-\frac{r_2}{2}\)^2-a^2}.
\end{split}\2.
\ee
According to Eq. \eqref{5}, the theoretical range of parameters within which internal and external event horizons exist can be determined as: $0\leqslant\frac{r_2}{M}\leqslant2(1-\frac{a}{M})$, or $-(1-\frac{r_2}{2M})\leqslant\frac{a}{M}\leqslant1-\frac{r_2}{2M}$. Since the spin parameter $a$ cannot exceed the black hole mass $M$, it can be deduced that the theoretical effective range for the parameter $r_2$ is $0\leqslant\frac{r_2}{M}\leqslant2$. In the next section, we will consider non-relativistic, steady-state, irrotational fluid accretion process in the spactime of Kerr-Sen black hole.

\section{Accretion solution for ultrarelativistic prefect fluid}
The primary objective of this section is to investigate the ultrarelativistic perfect fluid accretion solution. Specifically, the steady state fluid with an ultrarelativistic stiff equation of state which can be described as
\be
\label{6}
P=K\rho^2,
\ee
where $K$ is a constant, $\rho$ and $P$ are the proper mass density and the proper pressure, respectively. Since the perfect fluids satisfy the first law of thermodynamics: $\dif h=\frac{\dif P}{\rho}$ \cite{Font_2007} ($h$ represents the specific enthalpy defined as $\frac{u+P}{\rho}$ and $u$ denotes the total energy density), substituting this expression into the aforementioned formula, we obtain

\be
\label{7}
h=2K \rho.
\ee

In hydromechanics, the speed of sound plays a crucial role as a fundamental physical parameter for analyzing the velocity of fluid. We derive the fluid's sound speed  $c^2_s\equiv\sqrt{\frac{\p\ln{h}}{\p\ln{\rho}}}=1$ which indicates that the fluid's velocity is subsonic. The basic equations for investigating the evolution of the fluid are the continuity equation and the energy-momentum conservation equation, which have the following forms
\be
\1\{\begin{split}
\label{8}
&\na_{\m}J^{\m}=\na_{\m}\(\rho U^{\m}\)=0,\\
&\na_{\m}T^{\m\n}=\na_{\m}\(\rho hU^{\m}U^{\n}+P\delta^{\m}_{\n}\)=0.
\end{split}\2.
\ee
Then we can derive the Euler equation by combining the above two formulas with the first law
\be
\label{9}
U^{\m}\na_{\m}\(hU_{\n}\)+\na_{\n}h=0,
\ee
where $U^{\m}=\frac{\dif x^\m}{\dif \tau}$ is the four-velocity for fluid and satisfies $U^{\m}U_{\m}=-1$. The relativistic vorticity tensor is \cite{Petrich:1988zz}
\be
\label{10}
\om_{\m\n}=\na_{\n}\(hU_{\m}\)-\na_{\m}\(hU_{\n}\),
\ee
Utilizing the projection operator $P^{\mu}_{\nu}=\delta^{\mu}_{\nu}+U^{\mu}U_{\nu}$, we project the vorticity tensor into its spatial component.
\be
\label{11}
\tilde{\om}_{\m\n}=P^{\al}_{\m}P^{\bet}_{\n}\[\na_{\bet}\(hU_{\al}\)-\na_{\al}\(hU_{\bet}\)\].
\ee
Substituting \eqref{9} into the expression of \eqref{11}, one can demonstrate that for an irrotational fluid with a zero vortex tensor, $hU_{\mu}$ can be expressed as the gradient of the velocity potential $\Phi$.
\be
\label{12}
hU_{\m}=\na_{\m}\Phi,
\ee
Subsequently, the four-velocity normalization condition requires $h=\sqrt{-\na_{\m}\Phi\na^{\m}\Phi}$. Utilizing \eqref{12} and \eqref{8}, we obtain
\be
\label{13}
\na_{\m}\(\frac{\rho}{h}\na^{\m}\Phi\)=0,
\ee
it is a nonlinear differential equation and can be rewritten as
\be
\label{14}
\na_{\m}\na^{\m}\Phi=0.
\ee
The issue of thermodynamics is entirely transformed into the solution of a massless scalar field with boundary conditions. Nevertheless, it is essential to note that not all solutions to the equation are physically viable: a fluid's four-velocity must be required to be timelike. Within the permissible ranges of parameters, the theoretical values for the pressure and the specific enthalpy can be obtained by calculating the solution of the field $\Phi$.

\subsection{The solution in Kerr-Sen spacetime}
Petrich, Shapiro, and Teukolsky  investigated the solution of \eqref{14} in the Kerr spacetime under the assumption that the boundary conditions were fulfilled. In this subsection, we will follow their approach in the Kerr-Sen spacetime.

The analytical expression can be expressed as
\be
\label{15}
-\frac{\tilde{A}}{\tilde{\Delta}\tilde{\Sigma}}\p^2_{t}\Phi+\frac{1}{\tilde{\Sigma}}\p_r\(\tilde{\Delta}\p_r\Phi\)+\frac{1}{\tilde{\Sigma}\sin{\theta}}\p_{\theta}\(\sin{\theta}\p_{\theta}\Phi\)+\frac{\tilde{\Delta}-a^2\sin^2{\theta}}{\tilde{\Sigma}\tilde{\Delta}\sin^2{\theta}}\p_\phi^2\Phi-\frac{4Mra}{\tilde{\Delta}\tilde{\Sigma}}\p_t\p_\phi\Phi=0,
\ee
where $\tilde{A}$ is defined as
\be
\label{16}
\tilde{A}=\[r(r+r_2)+a^2\]^2-a^2\sin^2{\theta}\tilde{\Delta}.
\ee
According to the steady-state fluid accretion process, the solution is
\be
\label{17}
\Phi=e\[-t+\sum_{lm}R_{lm}(r)Y_{lm}(\theta,\phi) \],
\ee
where the velocity potential $\Phi$ has decomposed by the standard spherical harmonics which serve as a set of complete basis. The positive $e$ is related to the Bernoulli constant (per unit mass) and can be solved by
\be
\label{18}
e=-h U_{\m}\(\frac{\p}{\p t}\)^{\m}=-\p_t\Phi,
\ee
with the timelike Killing vector field $\(\frac{\p}{\p t}\)^\m=(1,0,0,0)$. Substituting \eqref{17} back into \eqref{15}, we can derive
\be
\label{19}
\frac{\dif}{\dif r}\[\tilde{\Delta}\p_r R_{lm}(r)\]-l(l+1)R_{lm}(r)+\frac{m^2a^2}{\tilde{\Delta}}R_{lm}(r)=0.
\ee
To simplify the radial component of \eqref{19}, we introduce a new variable $z$ to transforme the equation for $R_{lm}(r)$ into the Legendre equation.
\be
\1\{\begin{split}
\label{20}
&z\equiv\frac{r-M+\frac{r_2}{2}}{\sqrt{(M-\frac{r_2}{2})^2-a^2}},\\
&(1-z^2)\frac{\dif^2}{\dif z^2}R_{lm}(z)-2z\frac{\dif}{\dif z}R_{lm}(z)+l(l+1)R_{lm}(z)-\frac{(i\al m)^2}{1-z^2}R_{lm}(z)=0.
\end{split}\2.
\ee
Following this, Eq. \eqref{17} can be formulated in a general form.
\be
\label{21}
\Phi=e\[-t+\sum_{l}\(A_lP_l(z)+B_{l}Q_{l}(z)\)Y_{l0}+\sum_{lm}\(A^+_{lm}P^{im\al}_l(z)+A^-_{lm}P^{-im\al}_l(z)\)Y_{lm}(\theta,\phi)  \],
\ee
where $\al\equiv\frac{a}{\sqrt{\(M-\frac{r_2}{2}\)^2-a^2}}$, $m$ is a positive integer, and coefficients $A_l$, $B_{l}$, $A^+_{lm}$, and $A^-_{lm}$  could be determined by boundary conditions. $P_{l}(z)$, $P^{im\al}_{l}(z)$, and $Q_{l}(z)$ are Legendre functions of the first and the second kinds. $P^{im\al}_{l}(z)$ can be described as a hypergeometric function \cite{Babichev:2008dy},
\be
\1\{\begin{split}
\label{22}
&P^{im\al}_{l}(z)\propto \me^{im X}F\(-l,l+1,1-im\al;\frac{1-z}{2}\),  \\
&X\equiv\frac{\al}{2}\ln{\frac{z+1}{z-1}}=\frac{a}{2\sqrt{\(M-\frac{r_2}{2}\)^2-a^2}}\ln{\frac{r-r_{-}}{r-r_{+}}}.
\end{split}\2.
\ee
Using Eq. \eqref{12} and the normalization condition for the four-velocity, we obtain
\be
\1\{\begin{split}
\label{23}
&hU_t=-e,\\
&hU_r=\frac{e\[\sum_l\(A_l P'_l(z)+B_lQ'_l(z)\) Y_{l0}(\theta,\phi)+\sum_{lm}\(A^+_{lm}P'^{im\al}_l(z)+A^-_{lm}P'^{-im\al}_l(z)  \)Y_{lm}(\theta,\phi)\]}{\sqrt{(M-\frac{r_2}{2})^2-a^2}},\\
&hU_\theta=e\[\sum_{l}\(A_lP_l(z)+B_{l}Q_{l}(z)\)\frac{\p Y_{l0}(\theta,\phi)}{\p\theta}+\sum_{lm}\(A^+_{lm}P^{im\al}_l(z)+A^-_{lm}P^{-im\al}_l(z)\)\frac{\p Y_{lm}(\theta,\phi)}{\p\theta} \],\\
&hU_\phi=e\[\sum_{lm}\(A^+_{lm}P^{im\al}_l(z)+A^-_{lm}P^{-im\al}_l(z)\)\frac{\p Y_{lm}(\theta,\phi)}{\p\phi}\],\\
\end{split}\2.
\ee
with
\be
\label{24}
h^2=\frac{1}{\tilde{\Delta}\tilde{\Sigma}}\[e^2 \tilde{A}-\tilde{\Delta}^2(hU_r)^2-\tilde{\Delta}(hU_\theta)^2-\frac{\tilde{\Delta}-a^2\sin^2{\theta}}{\sin^2{\theta}}(hU_\phi)^2+4Mrae(hU_\phi)\].
\ee
The prime denotes the derivative with respect to $z$ and the analytical solution reveals a physical constraint: it must remain finite at $r_+$, necessitating $A^+_{lm}$ to be 0 (since $P^{im\alpha}_l$ is divergent at $r+$). Employing the limiting behavior as $z\rightarrow1$ (corresponding to $r\rightarrow r+$), we have
\be
\label{25}
h^2\rightarrow\frac{\[r_+(r_++r_2)+a^2\]^2e^2-\[\sqrt{(M-\frac{r_2}{2})^2-a^2}\sum_{l}B_{l}Y_{lm}(\theta,\phi) \]^2}{\tilde{\Delta}(r_+)\tilde{\Sigma}(r_+)}.
\ee
Analyzing the specific enthalpy with a focus on its continuity at the horizon, we conclude that only $B_0$ contributes, and the other terms $B_l (l>0)$ vanish. This implies that
\be
\label{26}
B_0=\frac{e[r_+(r_++r_2)+a^2]}{\sqrt{(M-\frac{r_2}{2})^2-a^2}},
\ee
where $Y_{00}$'s contribution is absorbed into $B_0$. By using the formula $Q_0(x)=\frac{1}{2}\ln{\frac{1+x}{1-x}}$, the associated solution degenerates to
\be
\1\{\begin{split}
\label{27}
&\Phi=e\[-t+\psi(r,\theta,\phi)\],\\
&\psi(r,\theta,\phi)\equiv\frac{r_+(r_++r_2)+a^2}{2\sqrt{(M-\frac{r_2}{2})^2-a^2}}\ln{\frac{r-r_-}{r-r_+}}+\sum_{l,m\geqslant0}A^-_{lm}\me^{-im\chi}F\(-l,l+1;1+im\al;\frac{1-z}{2}\)\\
&\times Y_{lm}(\theta,\phi).
\end{split}\2.
\ee
Since $\Phi$ is a real scalar field, imposing an constraint: $A^{-}_{l-m}=(-1)^m(A^{-}_{lm})^{*} $. In the study of steady-state accretion scenarios, it is common to assume an axisymmetric fluid distribution and a reflection symmetry about the equatorial plane. This allows us to focus only modes with $m=0$ and the angular quantum number $l$ must be an even integer. The coefficients $A^{-}_{lm}$ correspond to the distinctive structure of the fluid, and its determination will be addressed subsequently.

Eqs.\eqref{12}, \eqref{23} and \eqref{27} give the exact analytical formulas for $U^{\m}$ and the specific enthalpy, which yields
\be
\1\{\begin{split}
\label{28}
&\frac{hU^t}{e}=\frac{\tilde{A}}{\tilde{\Delta}\tilde{\Sigma}}-\frac{2Mra}{\tilde{\Delta}\tilde{\Sigma}}\p_\phi\psi,\\
&\frac{hU^r}{e}=\frac{\tilde{\Delta}}{\tilde{\Sigma}}\p_r\psi,\\
&\frac{hU^\theta}{e}=\frac{1}{\tilde{\Sigma}}\p_{\theta}\psi,\\
&\frac{hU^\phi}{e}=\frac{2Mra}{\tilde{\Delta}\tilde{\Sigma}}+\frac{\tilde{\Delta}-a^2\sin^2{\theta}}{\tilde{\Delta}\tilde{\Sigma}\sin^2{\theta}}\p_{\phi}\psi,\\
\end{split}\2.
\ee
and
\be
\label{29}
\frac{h^2}{e^2}=\frac{\tilde{A}}{\tilde{\Delta}\tilde{\Sigma}}-\frac{\tilde{\Delta}}{\tilde{\Sigma}}(\p_r\psi)^2-\frac{1}{\tilde{\Sigma}}(\p_\theta\psi)^2-\frac{\tilde{\Delta}-a^2\sin^2{\theta}}{\tilde{\Delta}\tilde{\Sigma}\sin^2{\theta}}(\p_{\phi}\psi)^2-\frac{4Mra}{\tilde{\Delta}\tilde{\Sigma}}(\p_\phi\psi).
\ee
There are two inherent constraints in \eqref{28} and \eqref{29}: the first constraint is that the right-hand side of Eq. (\ref{29}) must be positive, as the four-velocity $U^{\m}$ is required to be timelike. It is evident that not every point $r$ in Kerr-Sen spacetime satisfies this criteria.

Nevertheless, this issue can be addressed by selecting $A^{-}_{lm}$ small enough to ensure that $h$ is well-defined within the spherical shell's interval $r_+\leqslant r\leqslant R$ ($R$ is a spherical radius). The second constraint is that $U^r(r_+)<0$ for the fluid to flow radially into the event horizon. Fortunately, it can be demonstrated that this condition is automatically satisfied by the accretion solution of the ultrarelativistic fluid.

\subsection{Mass accretion rate for ultrarelativistic stiff fluid}

The accretion rate of a black hole is one of the most significant concept in astronomy. It is derived from the mass conservation flow and describes how rapidly the black hole absorbs surrounding fluids.
The constraint from stationary accretion is internally consistent with the following conditions: (1) the accreted matter is light-weight fluid, (2) the growth rate of black hole mass is slow \cite{Kumar:2017hgs, Malec:1999dd, Karkowski:2005zn}. Generally speaking, based on considerations by previous researchers on black hole accretion, it can be stated that these two conditions are generally satisfied due to the difficulty of the accreted matter's mass reaching the order of the black hole's mass. Therefore, when addressing the accretion of ordinary matter from the interstellar medium onto a sufficiently massive black hole, we can neglect the impact of back-reaction.

\begin{figure}[H]
\centering
\begin{minipage}{0.5\textwidth}
\centering
\includegraphics[scale=0.735,angle=0]{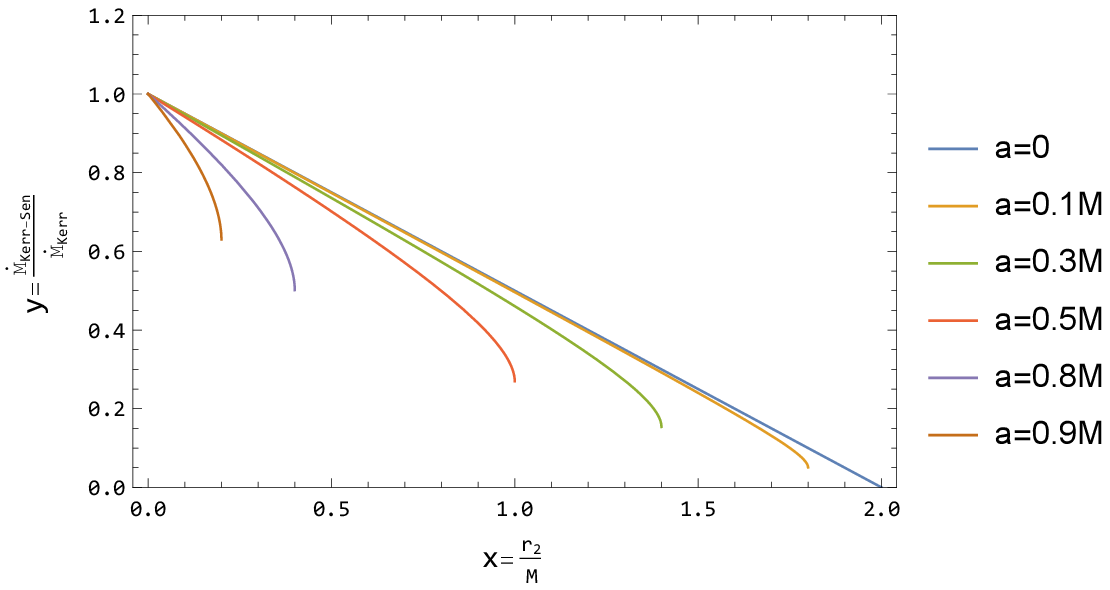}
\end{minipage}%
\begin{minipage}{0.5\textwidth}
\centering
\includegraphics[scale=0.735,angle=0]{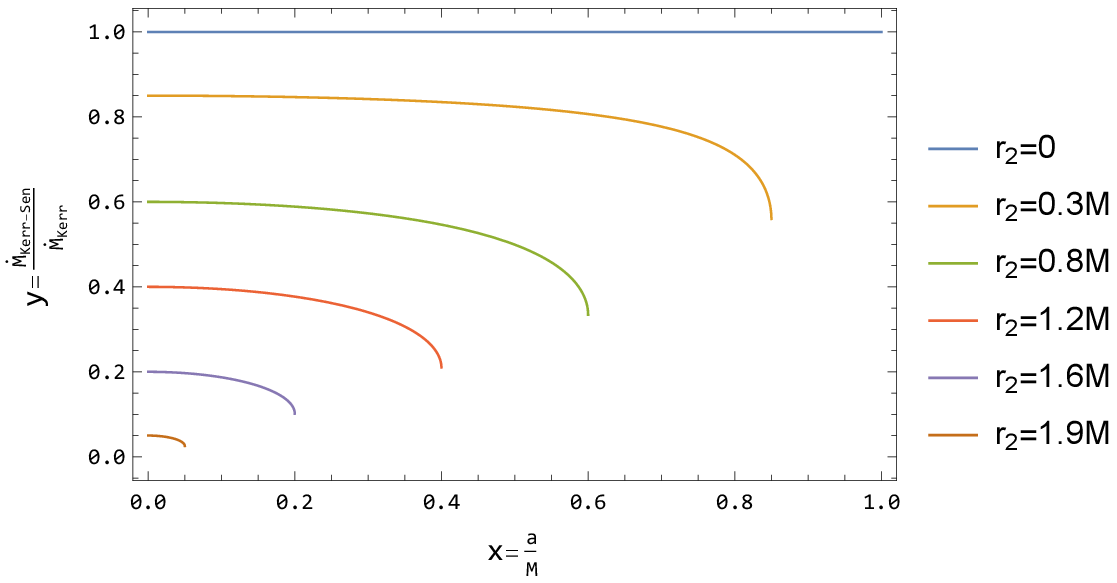}
\end{minipage}
\caption{\label{fig.1} The ratio of the accretion rate for the Kerr-Sen black hole to that of the Kerr black hole is illustrated with
a = 0, 0.1M, 0.3M, 0.5M, 0.8M, 0.9M and $r_2=$ 0, 0.3M, 0.8M, 1.2M, 1.6M, 1.9M.}
\end{figure}

The conserved quantities of  physical system are determined by the number of Killing vector fields. Given that the Kerr-Sen black hole exhibits symmetries in the $t$ and $\phi$ coordinates, we can define the conservation flow
\be
\1\{\begin{split}
\label{30}
&J^{\m}_{\varepsilon}=-T^{\m}_{\n}\(\frac{\p}{\p t}\)^{\n},\\
&J^{\m}_{L}=T^{\m}_{\n}\(\frac{\p}{\p \phi}\)^{\n},
\end{split}\2.
\ee
which correspond to the energy, the angular momentum, and the mass flow $J^{\m}=\rho U^{\m}$, respectively. Combining \eqref{12} with the stiff equation of state yields the deduced conserved current as follows
\be
\1\{\begin{split}
\label{31}
&J^{\m}=\frac{\rho}{h}\na^{\m}\Phi,\\
&T^{\m}_{\n}=\frac{\rho}{h}\(\na^{\m}\Phi\na_{\n}\Phi-\frac{1}{2}\delta^{\m}_{\n}\na_{\sig}\Phi\na^{\sig}\Phi\).
\end{split}\2.
\ee
The mass flow through the surface of a sphere with a certain radius is referred to the mass accretion rate, which yields
\be
\label{32}
\dot{\mathscr{M}}=-\int_{S}J^r\sqrt{-g}\dif \theta\dif\phi,
\ee
with dot denoting the time derivative and $S$ meaning any sphere of radius $r$.
In particular, it should be emphasized that $\dot{\mathscr{M}}$ is independent of the position chosen during the steady-state accretion process. Therefore, we select the surface $r=r_+$.
In order to have a explicit expression for the accretion rate, we solve $\dot{\mathscr{M}}$ by substituting  $J^r=\frac{\rho e\tilde{\Delta}}{\Sigma}\p_r\psi$  and \eqref{27} into \eqref{32},
\be
\label{33}
\dot{\mathscr{M}}=-\frac{\rho e}{h}\int_{r=r_+}\tilde{\Delta}\p_r\psi\sin{\theta}\dif\theta\dif\phi=\frac{4\pi\rho e}{h}\[r_+(r_++r_2)+a^2\].
\ee
It appears that the ultrarelativistic fluid with $\frac{\rho}{h}=\frac{1}{2K}$ can be moved outside of the integral. The energy accretion rate $\dot{M}$ and angular momentum accretion rate $\dot{J}$ could be expressed as
\be
\1\{\begin{split}
\label{34}
&\dot{M}=-\int_{r=r_+}J^r_{\varepsilon}\tilde{\Sigma}\sin{\theta}\dif\theta\dif\phi=e\dot{\mathscr{M}},\\
&\dot{J}=-\int_{r=r_+}J^r_{L}\tilde{\Sigma}\sin{\theta}\dif\theta\dif\phi=0.
\end{split}\2.
\ee
The formulas indicate that both the mass accretion rate and energy accretion rate remain constant. The transformation of angular momentum is zero, which implies that the steady-state fluid is axisymmetric $(m=0)$ in the Boyer-Lindquist coordinate system. The ratio of the accretion rate for the Kerr-Sen black hole to that of the Kerr black hole is depicted in Figure \ref{fig.1}. Both Figures indicate that within the theoretically permissible range of parameters, as $a$ and $r_2$ increase, the accretion rate of the Kerr-Sen black hole decreases significantly, comparing with that of the Kerr black hole.

\section{The quadrupolar flow Solution in ZAMO Framework}
To investigate the relative velocity in three dimensions, it is more convenient to adpot the zero angular momentum observer (ZAMO) framework \cite{Frolov_2014,Liao:2016mum}. The observer's four-velocity depends on $\frac{\p}{\p t}+\Omega\frac{\p}{\p \phi}$ ($\Omega=\frac{2Mar}{\tilde{A}}$ is the angular velocity). On the background of the Kerr-Sen black hole, the four orthogonal basis vectors could be expressed as
\be
\1\{\begin{split}
\label{35}
&e_{\hat{t}}=\sqrt{\frac{\tilde{A}}{\tilde{\Sigma}\tilde{\Delta}}}\(1,0,0, \Omega \),
~e_{\hat{r}}=\sqrt{\frac{\tilde{\Delta}}{\tilde{\Sigma}}}\(0,1,0,0\),\\
&e_{\hat{\theta}}=\frac{1}{\sqrt{\tilde{\Sigma}}}\(0,0,1,0\),
~~~~e_{\hat{\phi}}=\sqrt{\frac{\tilde{\Sigma}}{\tilde{A}\sin^2{\theta}}}\(0,0,0,1\).
\end{split}\2.
\ee
According to the coordinate transformational relation $U^{\hat{\m}}=e^{\hat{\m}}_{\beta}U^{\beta}$, the four-velocity within ZAMO framework are
\be
\1\{\begin{split}
\label{36}
&\frac{h}{e}U^{\hat{t}}=\sqrt{\frac{\tilde{A}}{\tilde{\Sigma}\tilde{\Delta}}}\(1-\Omega\p_{\phi}\psi \),
~\frac{h}{e}U^{\hat{r}}=\sqrt{\frac{\tilde{\Delta}}{\tilde{\Sigma}}}\p_{r}\psi,\\
&\frac{h}{e}U^{\hat{\theta}}=\frac{1}{\sqrt{\tilde{\Sigma}}}\p_{\theta}\psi,
~~~~~~~~~~~~~~\frac{h}{e}U^{\hat{\phi}}=\sqrt{\frac{\tilde{\Sigma}}{\tilde{A}\sin^2{\theta}}}\p_{\phi}\psi.\\
\end{split}\2.
\ee
The Lorentz factor is $\Gamma\equiv U^{\hat{t}}=\frac{1}{\sqrt{1-V^2}}$, and the additional three-velocity corresponding to Eq. \eqref{36} can be defined as
\be
\1\{\begin{split}
\label{37}
&V^{\hat{r}}=\frac{U^{\hat{r}}}{U^{\hat{t}}}=\frac{\tilde{\Delta}\p_r\psi}{\sqrt{\tilde{A}}\(1-\Omega\p_{\phi}\psi \)},\\
&V^{\hat{\theta}}=\frac{U^{\hat{\theta}}}{U^{\hat{t}}}=\frac{\sqrt{\tilde{\Delta}}\p_{\theta}\psi}{\sqrt{\tilde{A}}\(1-\Omega\p_{\phi}\psi \)},\\
&V^{\hat{\phi}}=\frac{U^{\hat{\phi}}}{U^{\hat{t}}}=\frac{\tilde{\Sigma}\sqrt{\tilde{\Delta}}\p_\phi\psi}{\tilde{A}\sin{\theta}\(1-\Omega\p_{\phi}\psi \)},\\
\end{split}\2.
\ee
with
\be
\label{38}
V=\sqrt{(V^{\hat{r}})^2+(V^{\hat{\theta}})^2+(V^{\hat{\phi}})^2}.
\ee
Evidently, these representations of three- and four-velocities lead to variety of relevant conclusions. The timelike characteristic of four velocity requires $V<1$ ($U^{\hat{t}}>0$). Therefore, we can constrain the three-velocity highlighted above according to the following two conditions.

\be
\label{39}
\Omega\p_{\phi}\psi<1,
\ee
and
\be
\label{40}
V^2=\frac{1}{\tilde{A}\(1-\Omega\p_{\phi}\psi \)^2}\[(\tilde{\Delta}\p_{r}\psi)^2+\tilde{\Delta}(\p_{\theta}\psi)^2+\frac{\tilde{\Delta}\tilde{\Sigma}^2}{\tilde{A}\sin^2{\theta}}(\p_{\phi}\psi)^2\]<1\\.
\ee
The case of $(l,m)=(2,0)$, which is the essential solution utilized to characterize choked accretion, also constitutes the solution for the axisymmetric quadrupolar flow. This solution inherently satisfies the constraint equation \eqref{39} ($\partial_{\phi}\psi=0$). The restriction imposed by \eqref{40} will be explained in more detail later. Specifically, the Bondi-Michel-type accretion corresponds to $l=0$  and  the wind accretion  was accompanied by $l=1$, both extensively explored in \cite{1993AmJPh..61..825K,Tejeda:2017xqt}. We extend hypergeometric series near the horizon and obtain quadrupolar flow's velocity potential $\Phi$
\be
\1\{\begin{split}
\label{41}
&\Phi=e\[-t+\frac{r_+\(r_++r_2\)+a^2}{2\sqrt{\(M-\frac{r_2}{2}\)^2-a^2}}\ln{\frac{r-r_-}{r-r_+}}+NF\(r,\theta,\phi\)\],\\
&F\(r,\theta,\phi\)=(3\cos^2{\theta}-1)\(3r^2-6Mr+2M^2+a^2-2Mr_2+3rr_2+\frac{1}{2}r^2_2\),
\end{split}\2.
\ee
where $N$ represents the coefficient $A^-_{20}$. Inserting Eq. \eqref{41} into Eqs. \eqref{36} and \eqref{37}, we can derive
\be
\1\{\begin{split}
\label{42}
&V^{\hat{r}}=\frac{\tilde{\Delta}N F_{,r}-2Mr_+}{\sqrt{\tilde{A}}},\\
&V^{\hat{\theta}}=\sqrt{\frac{\tilde{\Delta}}{\tilde{A}}}  N F_{,\theta},\\
&V^{\hat{\phi}}=0,\\
\end{split}\2.
\ee
and
\be
\label{43}
\frac{h^2}{e^2}=\frac{\tilde{A}(1-V^2)}{\tilde{\Sigma}\tilde{\Delta}}=\frac{1}{\tilde{\Sigma}}\[\frac{\tilde{A}}{\tilde{\Delta}}-\tilde{\Delta}\(NF_{,r}-\frac{2Mr_+}{\tilde{\Delta}}\)^2-N^2 F^2_{,\theta}\],
\ee
with
\be
\1\{\begin{split}
\label{44}
&F_{,r}=(3\cos^2{\theta}-1)\(6r-6M+3r_2\),\\
&F_{,\theta}=-6\cos{\theta}\sin{\theta}\(3r^2-6Mr+2M^2+a^2-2Mr_2+3rr_2+\frac{1}{2}r^2_2\).\\
\end{split}\2.
\ee
The fluid distribution is entirely defined by Eq. \eqref{42}. By resolving the stagnation point, we can figure out its morphological structure. The stagnation related to $V^{\hat{\theta}}=0$ could be resolved with $\theta=0$, $\frac{\pi}{2}$, and $\pi$, while $V^{\hat{r}}=0$ will ultimately determine the location of the stagnation point. Since the flow is symmetric in the equatorial plane, there are two scenarios.

Case 1 $(\theta=0,\pi)$: $N>0$ with inflow traveling through the equatorial plane $(\theta=\frac{\pi}{2})$ and outflow along the polar axis. On the polar axis, the stagnation point $r=r_s$ is symmetrical distribution and fulfills
\be
\label{45}
N=\frac{Mr_+}{(6r_s-6M+3r_2)(r_s-r_+)(r_s-r_-)}.
\ee

\begin{figure}[H]
\centering
\begin{minipage}{0.5\textwidth}
\centering
\includegraphics[scale=0.7,angle=0]{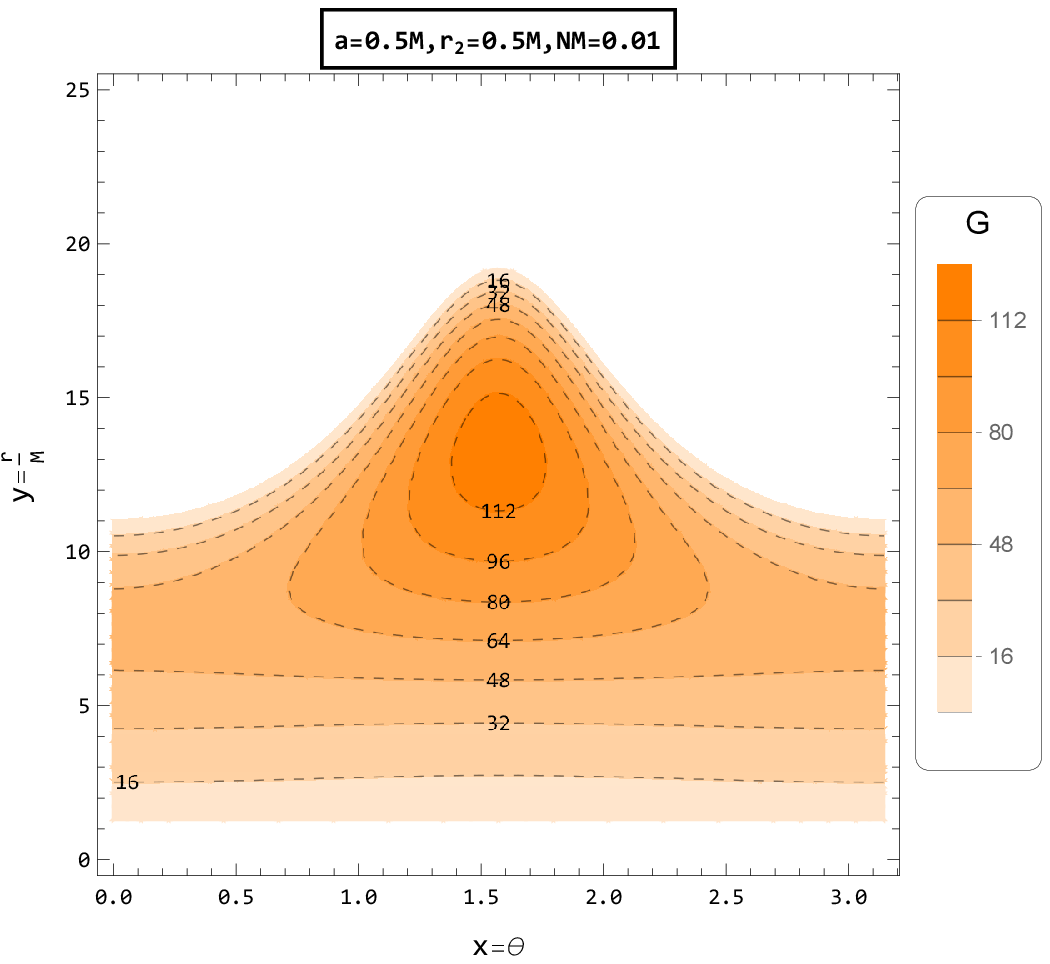}
\end{minipage}%
\begin{minipage}{0.5\textwidth}
\centering
\includegraphics[scale=0.7,angle=0]{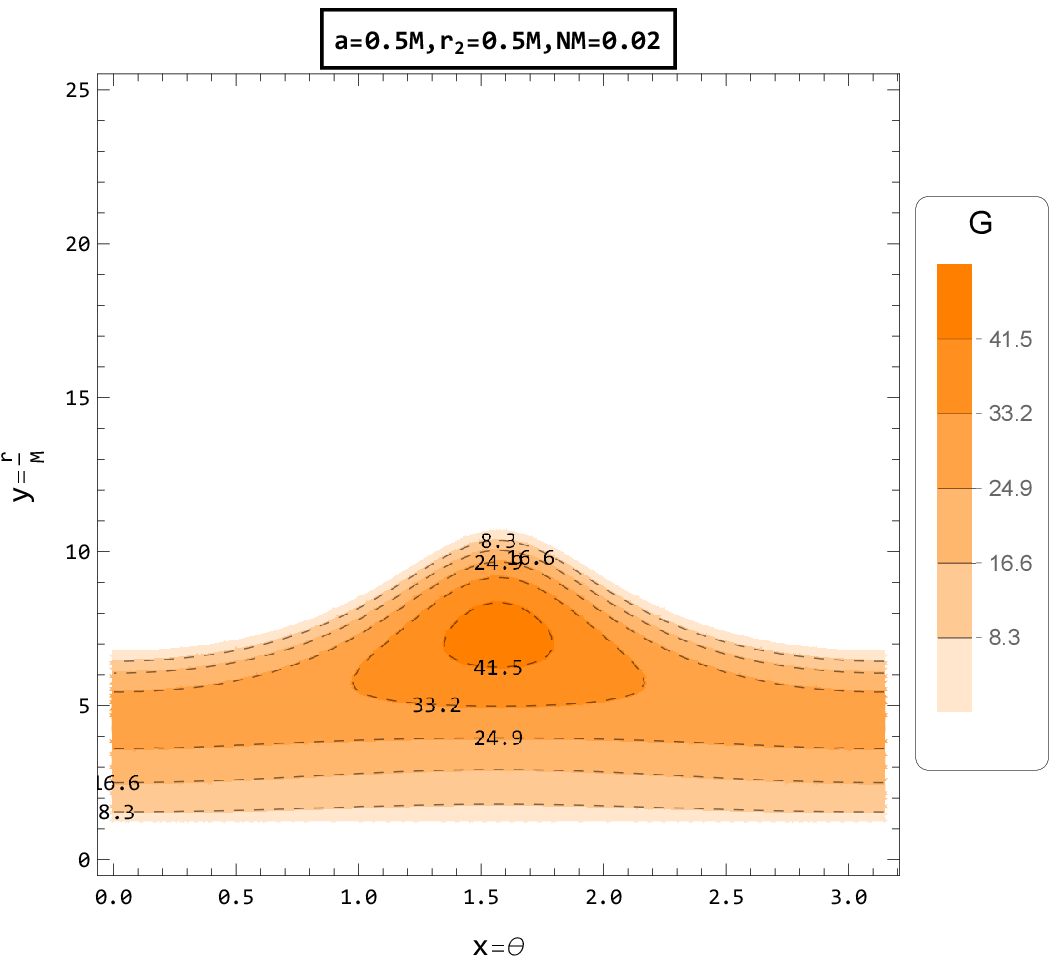}
\end{minipage}
\caption{\label{fig.2} The above contour maps represent $G$ with $a=0.5M$, $r_2=0.5M$, some values of $N$, and $NM$ = 0.01, 0.02, respectively. }
\end{figure}

Case 2 $(\theta=\frac{\pi}{2})$: $N<0$ is associated with inflow entering at both ends of the polar axis and outflow along the equatorial plane. In this case the stagnation point exists symmetrically in the equatorial plane.
\be
\label{46}
N=-\frac{2Mr_+}{(6r_s-6M+3r_2)(r_s-r_+)(r_s-r_-)}.
\ee
When the black hole's parameters remain constant, the coefficients in the potential function entirely determines the location of the stationary point. In quadrupolar flow situation, the constraint in \eqref{40} is equivalent to the positivity of the right-hand side of the last term in \eqref{43}. It follows that
\be
\label{47}
G\equiv g_0(r)\cos^4{\theta}+g_1(r)\cos^2{\theta}+g_2(r)>0,
\ee
with
\be
\1\{\begin{split}
\label{48}
&g_0(r)=9N^2\(2a^2+4M^2+6r^2+6rr_2+r^2_2-4M(3r+r_2)\)^2-9N^2(6r-6M+3r_2)^2\tilde{\Delta},\\
&g_1(r)=12NMr_+(6r-6M+3r_2)+6N^2(6r-6M+3r_2)^2\tilde{\Delta}+a^2\\
&-9N^2\(2a^2+4M^2+6r^2+6rr_2+r^2_2-4M(3r+r_2)\)^2,\\
&g_2(r)=-N^2(6r-6M+3r_2)^2\tilde{\Delta}-4NMr_+(6r-6M+3r_2)+r^2
+rr_2+2Mr+4M^2\frac{r+r_+}{r-r_-}.\\
\end{split}\2.
\ee
As the streamline is minimally influenced by the black hole parameter $a$, we fix $a=0.5M$ (corresponding to the constrained range: $0<\frac{r_2}{M}<1$) and examine how the coefficient $N$ affects the streamline.
The physical solution relies on the four-velocity's timelike restrictions shown in the isothermal diagram Figure \ref{fig.2}. The horizontal axis indicates angle $\theta$ and the vertical axis represents the dimensionless value of $r$. We observe that the coefficient $N$ has an impact within the confined physical regions of parameters. The diagram becomes narrower as $N$ increases, until there is no longer a desirable region. Therefore, the value of $N$ must be small enough to guarantee that the requirement for the fluid being timelike is satisfied within a sufficiently large radius $R$.

\begin{figure}[H]
\centering
\begin{minipage}{0.5\textwidth}
\centering
\includegraphics[scale=0.6,angle=0]{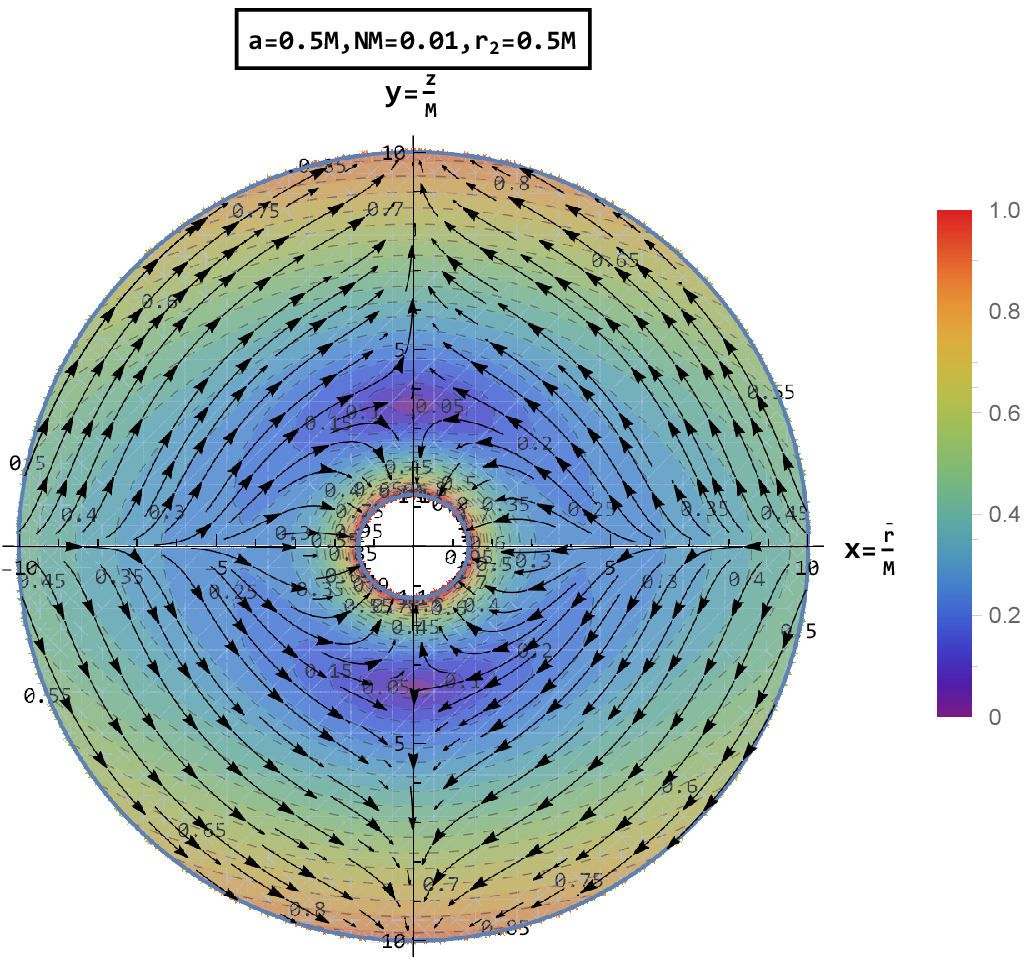}
\end{minipage}%
\begin{minipage}{0.5\textwidth}
\centering
\includegraphics[scale=0.6,angle=0]{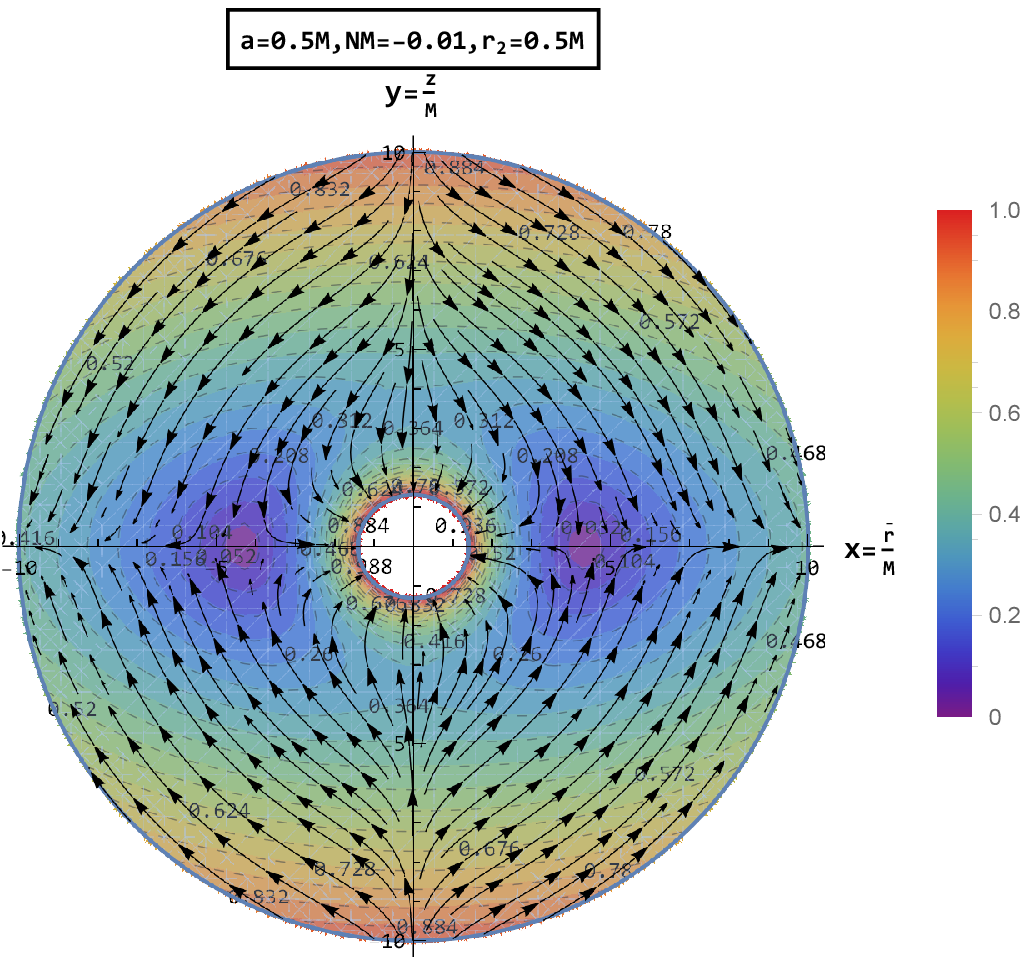}
\end{minipage}
\caption{\label{fig.3} Two figures correspond to the streamline (solid arrow) and the isocontour (dotted line). The magnitude of the three-velocity is entirely determined by the change in color. The diagram use cylindrical-like coordinate transformation: $\overline{r}=\sqrt{r^2+a^2}\sin{\theta}$, $z=r\cos{\theta}$.}
\end{figure}
The streamline and isocontour diagram of the non-relativistic stiff fluid determined by \eqref{42} are shown in Figure \ref{fig.3}. The isocontour of three-velocity are represented by the dashed line, the streamline is shown by the black arrow, and the stagnation point is symbolized by the blue area. The diagram illustrates that part of the fluid flows into the black hole, while other parts flow towards the poles or the equatorial plane. We will select $R=10M$ as the outer radius for investigation in the following discussions, since the timelike characteristic of velocity is assured in the interval $(r_+, R)$. It is worth noticing that the black hole domain is symbolised by the hole in the middle, which also highlights that $\Phi$ is not well defined in the ZAMO frame.

\section{Choked accretion}
The choked accretion is a hydrodynamic mechanism that has been applied to the investigations of accretion onto Schwarzschild or Kerr black holes \cite{Aguayo-Ortiz:2020qro,Tejeda:2019fwr,Aguayo-Ortiz:2019fap}. The mechanism describes that the fluid is injected radially along the equatorial plane. If the injection velocity is excessively high, the anisotropic density distribution fluctuates dramatically, leading to the deviation of part of the fluid from its original orbit and subsequent ejection along the poles. This process is reversible, as seen in cases 1 and 2, respectively. We'll focus on $N>0$ (case 1) in this section.

It was mentioned previously that to fulfill the timelike requirements for four-velocity a spherical surface with a radius of $R$ must be supplied. Physical values within the spherical shell are well-defined. Simultaneously, it is crucial to establish a reference point for measuring physical quantities. The injection rate $V_0$ is set at $\(R,\frac{\pi}{2}\)$ in equatorial plane and ejection rate $V_{ej}$ is determined at $(R,0)$, both can be represented as

\be
\1\{\begin{split}
\label{49}
&V_0=V^{\hat{r}}\(R,\frac{\pi}{2}\)=\frac{\tilde{\Delta}_0\(6R-6M+3r_2\)N+2Mr_+}{\sqrt{\tilde{A_0}}},  \\
&V_{ej}=V^{\hat{r}}\(R,0\)=\frac{2V_0\sqrt{\tilde{A_0}}-6Mr_+}{R(R+r_2)+a^2}, \\
\end{split}\2.
\ee

with
\be
\1\{\begin{split}
\label{50}
&\tilde{\Delta}_0=(R-r_+)(R-r_-),\\
&\tilde{A}_0=R(R+r_2)(a^2+R^2+Rr_2)+2MRa^2.\\
\end{split}\2.
\ee
Since the fluid is subluminal and it flows out along both ends of the polar axis, we must require $(0<V_{ej}<1)$ to acquire the range of the initial velocity $V_0$ as
\be
\label{51}
\frac{3Mr_+}{\sqrt{\tilde{A}_0}}<V_0<\frac{R(R+r_2)+a^2+6Mr_+}{2\sqrt{\tilde{A}_0}}.
\ee
Up to this point, we haven't determined the stagnation point or the value of the coefficient $N$ in principle. Therefore, boundary constraints must be imposed. It can be concluded from \eqref{37} that
\be
\label{52}
e=h\Gamma\sqrt{\frac{\tilde{\Sigma}\tilde{\Delta}}{\tilde{A}}}=h_0\Gamma_0\sqrt{\frac{\tilde{\Sigma}_0\tilde{\Delta}_0}{\tilde{A_0}}},
\ee
where
\be
\1\{\begin{split}
\label{53}
&\Gamma_0=\frac{1}{\sqrt{1-V^2_0}},\\
&\tilde{\Sigma}_0=R(R+r_2).\\
\end{split}\2.
\ee
Using the \eqref{7} and substituting \eqref{52} into \eqref{37}, we can derive
\be
\label{54}
\frac{h}{h_0}=\frac{\rho}{\rho_0}=\frac{\Gamma_0\sqrt{\tilde{\Sigma}_0\tilde{\Delta}_0\tilde{A}}}{\Gamma\sqrt{\tilde{\Sigma}\tilde{\Delta}\tilde{A}_0}},
\ee
where the density $\rho_0=\rho\(R,\frac{\pi}{2}\)$ and the specific enthalpy $h_0=h\(R,\frac{\pi}{2}\)$ are determined by the parameters $M$, $a$ and $r_2$ with values at the reference point. According to \eqref{49}, it follows that
\be
\label{55}
N=\frac{V_0\sqrt{\tilde{A}}_0-2Mr_+}{\tilde{\Delta}_0(6R-6M+3r_2)},
\ee
Simultaneously, the two equivalent conditions \eqref{45} and \eqref{55} give the location of stagnation point
\be
\label{56}
r_s=M-\frac{r_2}{2}+\(\lam-\sqrt{\lam^2-\frac{\(M^2-a^2-Mr_2+\frac{1}{4}r^2_2\)^3}{27}}\)^{\frac{1}{3}}+\(\lam+\sqrt{\lam^2-\frac{\(M^2-a^2-Mr_2+\frac{1}{4}r^2_2\)^3}{27}}\)^{\frac{1}{3}},
\ee
with
\be
\label{57}
\lam\equiv\frac{Mr_+(R-r_+)(R-r_-)(R-M+\frac{r_2}{2})}{2(V_0\sqrt{\tilde{A_0}}-2Mr_+)}.
\ee
The stagnation point, the black hole event horizon, the ratio of accretion rate and the ejection rate are presented in the Table \ref{table.1}. The initial column in the table delineates the scenario of the Kerr black hole. The tabulated data demonstrate that the dilaton parameter $r_2$ increases within the range allowed by the model and a reduction is observed in the location of the stagnation point. In contrast to this pattern, the jet ejection rate exhibits an augmentation with escalating value of the parameter.

\begin{table}[H]
\centering
\renewcommand\arraystretch{1}{
\setlength{\tabcolsep}{2mm}{
\begin{tabular}{|c|cccccccccc|}
\hline
&\multicolumn{10}{|c|}{$a=0.5$ M, $R=10$ M, $V_0=0.4$} \\
\hline
$\frac{r_2}{M}$  &0& 0.1& 0.2 &0.3& 0.4& 0.5& 0.6& 0.7& 0.8& 0.9 \\
\hline
$\frac{r_s}{M}$& 4.410& 4.287& 4.160& 4.030 &3.895& 3.754 &3.605 &3.444& 3.267 &3.056\\
\hline
$\frac{r_{+}}{M}$&1.866& 1.758&1.648&1.537& 1.425& 1.310&1.190& 1.065& 0.932&0.780\\
\hline
$\frac{\dot{\mathscr{M}}_{Kerr-Sen}}{\dot{\mathscr{M}}_{Kerr}}$   &1.000& 0.942& 0.883& 0.824&0.763& 0.702& 0.638& 0.571&0.499&0.418\\
\hline
$V_{ej}$     &  0.687& 0.695& 0.702&0.710& 0.717& 0.725& 0.732& 0.740& 0.748& 0.756\\
\hline
$\delta_{\rho}$    & 0.479& 0.492& 0.504&0.517&0.529& 0.541&0.554& 0.566& 0.579&0.594\\
\hline
\end{tabular}}}
\caption{\label{table.1}The location of stagnation point, the black hole event horizon, the ratio of accretion rate, the relative density ratio, and the ejection rate are shown in the table with $a=0.5$ M, $R=10$ M, $V_0=0.4$ .            }
\end{table}

\begin{figure}[H]
\centering
\begin{minipage}{0.5\textwidth}
\centering
\includegraphics[scale=0.7,angle=0.0]{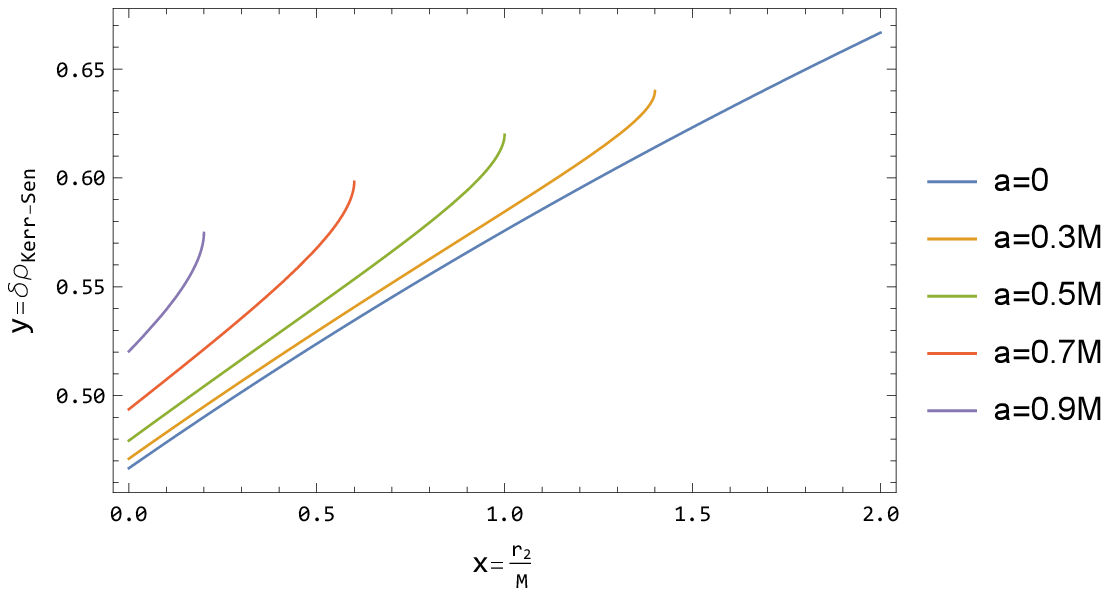}
\end{minipage}%
\begin{minipage}{0.5\textwidth}
\centering
\includegraphics[scale=0.7,angle=0.0]{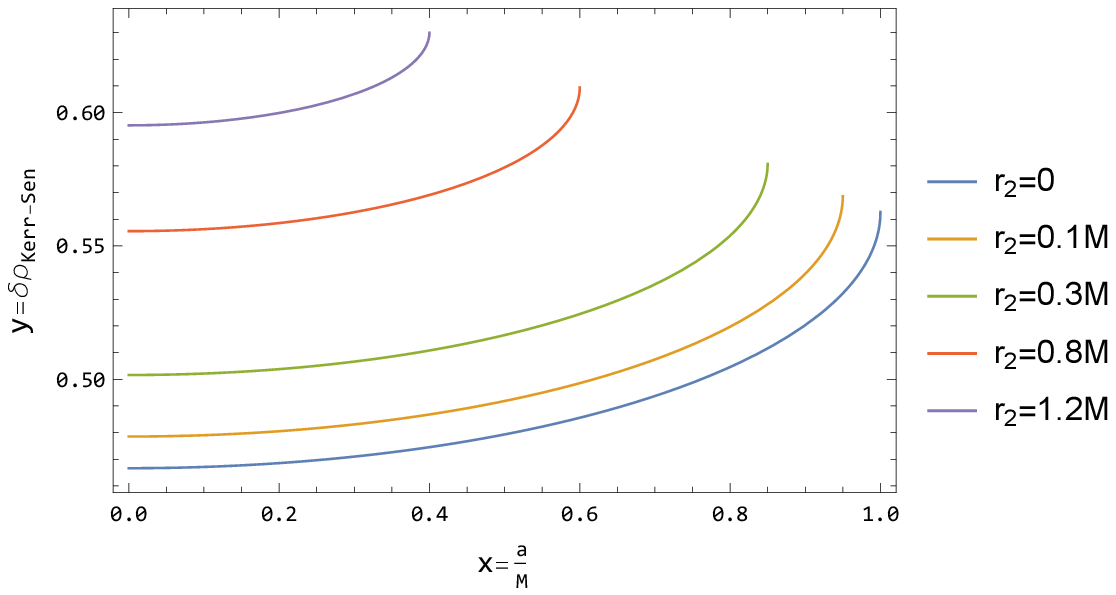}
\end{minipage}%
\caption{\label{fig.4}The relative density ratio as a function of $r_2/M$ (the left figure) or $a/M$ (the right figure) with $V_0=0.4$, $R=10$ M and some values of the parameter $a$ or $r_2$.}
\end{figure}

\subsection{Distribution of the density and the streamlines }
 With Eq. \eqref{54}, we define a quantity $\delta$ that describes the relative deviation of the density taking the ejection's value at $\(R, 0\)$ from taking the injection's value at $\(R, \frac{\pi}{2}\)$, which yield
\be
\label{58}
\delta_\rho\equiv1-\frac{\rho\(R,0\)}{\rho\(R,\frac{\pi}{2}\)}=1-\sqrt{\frac{\(1-V^2_{ej}\)R(R+r_2)\(R(R+r_2)+a^2\)}{\(1-V^2_0\)\(R(R+r_2)(a^2+R^2+Rr_2)+2MRa^2\)}}.
\ee
We observe that the density of location $(R,0)$ tends to zero for $\delta_\rho=1$, while the density of the ejection point is equivalent to injection point for $\delta_\rho=0$.
The inferences derived from the final row of Table \ref{table.1} and Figure \ref{fig.4} indicate that in the black hole an increseasing parameter $r_2$ leads to an increased relative density ratio, while a lower density ratio for the Kerr-Sen black hole, compared to the Kerr black hole.

\begin{figure}[H]
\centering
\begin{minipage}{0.5\textwidth}
\centering
\includegraphics[scale=0.6,angle=0]{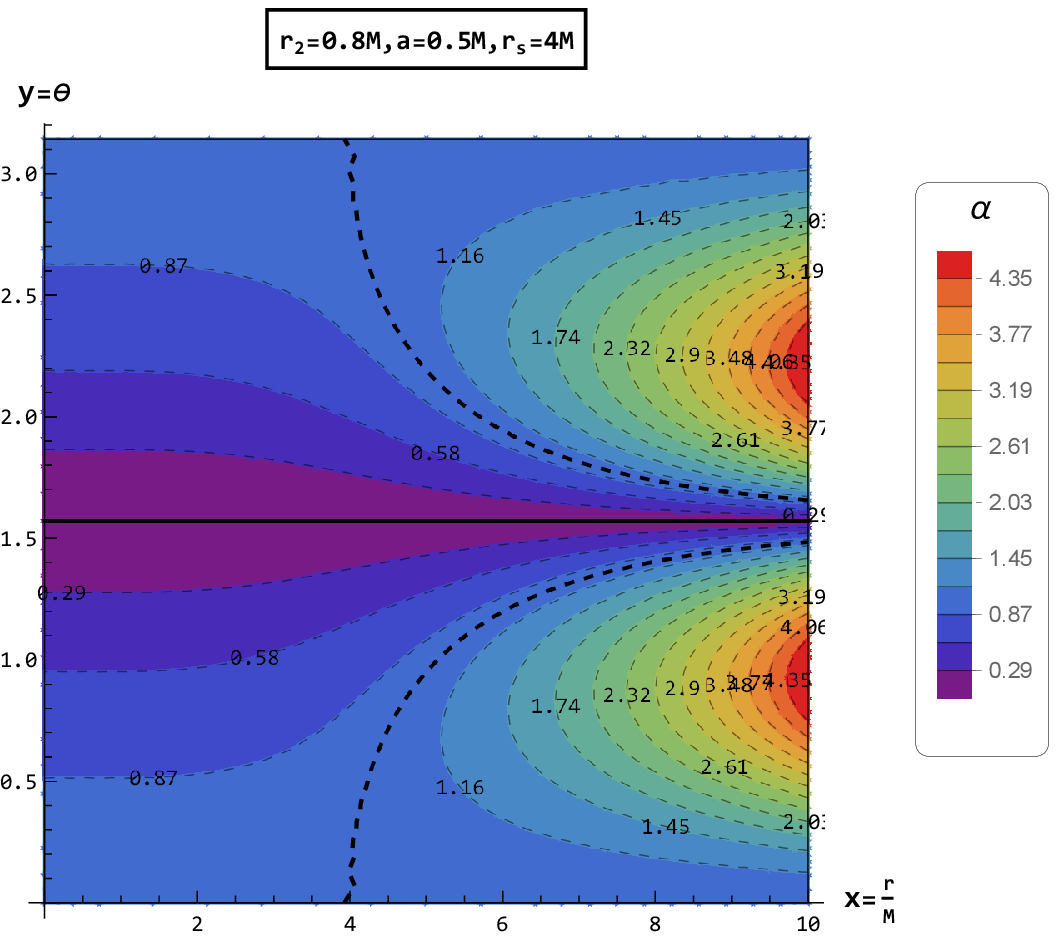}
\end{minipage}%
\begin{minipage}{0.5\textwidth}
\centering
\includegraphics[scale=0.6,angle=0]{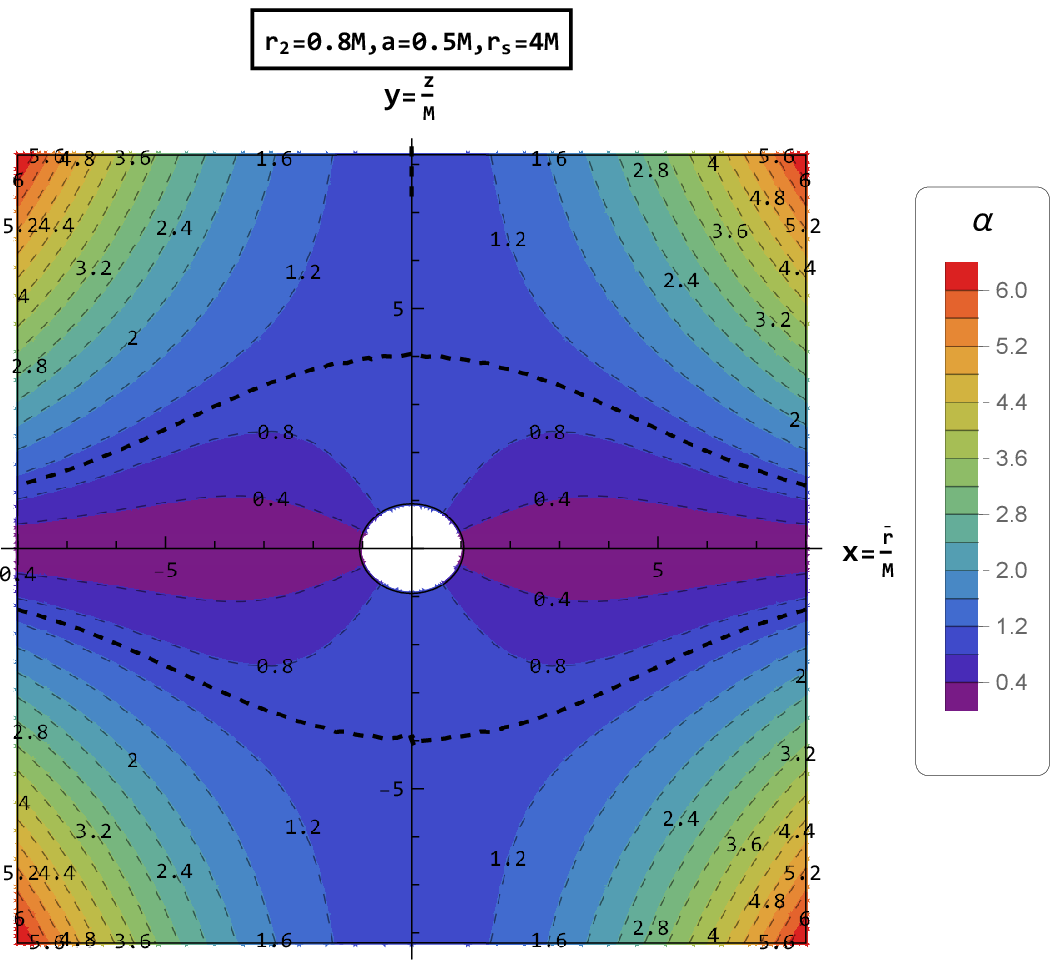}
\end{minipage}
\caption{\label{fig.5} For two figures: $r_2=0.8$ M, $a=0.5$ M, and $r_s=4$ M . The solid line in the left figure denotes the line with $\theta=\frac{\pi}{2}$, while the isocontour lines are represented by dashed lines. The coordinate transformations, $\overline{r}=\sqrt{r^2+a^2}\sin{\theta}$ and $z=r\cos{\theta}$ are adopted in the right figure, and the centre empty region represents the interior of Kerr-Sen black hole. }
\end{figure}

Eventually, we obtain a streamlined diagram which can depict the trajectories of fluid. Applying Eqs. \eqref{42} to eliminate time $t$, we derive
\be
\label{59}
\al=\cos{\theta} \[1+\frac{\sin^2{\theta}\(r-r_+\)\(r-r_-\)\(r-M+\frac{r_2}{2}\)}{2\(r_s-r_+\)\(r_s-r_-\)\(r_s-M+\frac{r_2}{2}\)}\],
\ee
where $\al$ is the constant of integration. Streamlines with $|{\al}|=1$ are connected to the stagnation point, whereas those with $|\al|>1$ escape via the bipolar outflow, and those with $|\al|<1$ are absorbed into black hole. The streamlines are described in Figure \ref{fig.5}. In the left diagram, we observe that the whole diagram is symmetrically distributed along the line $\theta=\frac{\pi}{2}$ (the middle solid line); that the dashed lines depict the trajectories of streamline which differs from each other with different values of $|{\al}|$. It is straightforward to notice that the dividing line ($|{\al}|=1$) indicates whether the fluid is absorbed into or ejected from the black hole. The right streamlined diagram is shown in the cylindrical-like coordinate system ($\overline{r},z$). Similarly, the boundary of the blue region is the line with $|{\al}|=1$ denoted by the dotted line.

\subsection{Injection rate, ejection rate, and critical angle}
\begin{figure}[H]
\centering

\begin{minipage}{0.5\textwidth}
\centering
\includegraphics[scale=0.7,angle=0]{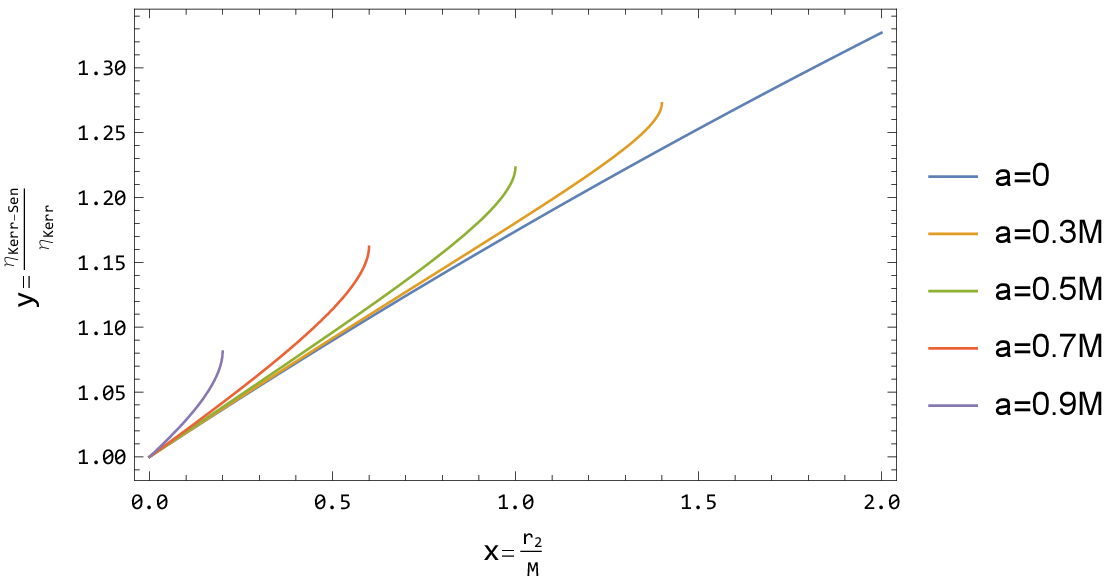}
\end{minipage}%
\begin{minipage}{0.5\textwidth}
\centering
\includegraphics[scale=0.7,angle=0]{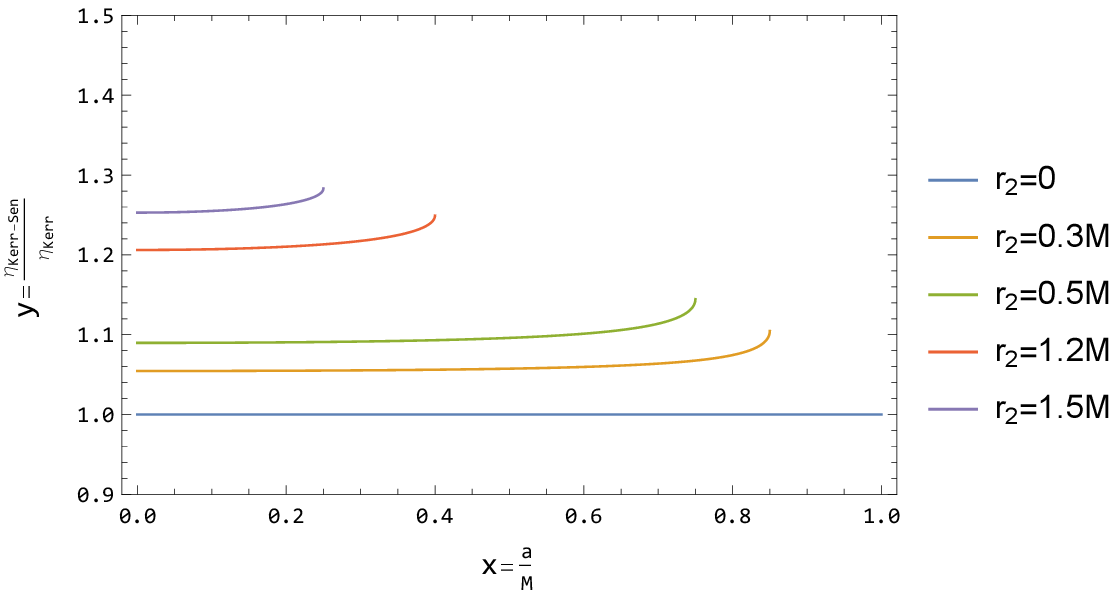}
\end{minipage}
\begin{minipage}{0.5\textwidth}
\centering
\includegraphics[scale=0.7,angle=0]{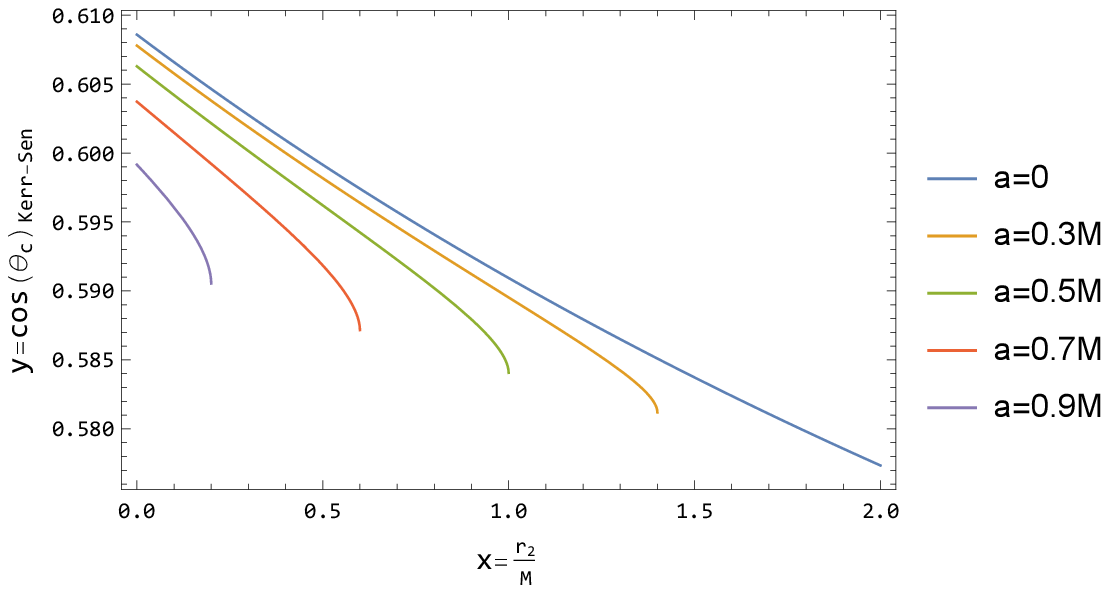}
\end{minipage}%
\begin{minipage}{0.5\textwidth}
\centering
\includegraphics[scale=0.7,angle=0]{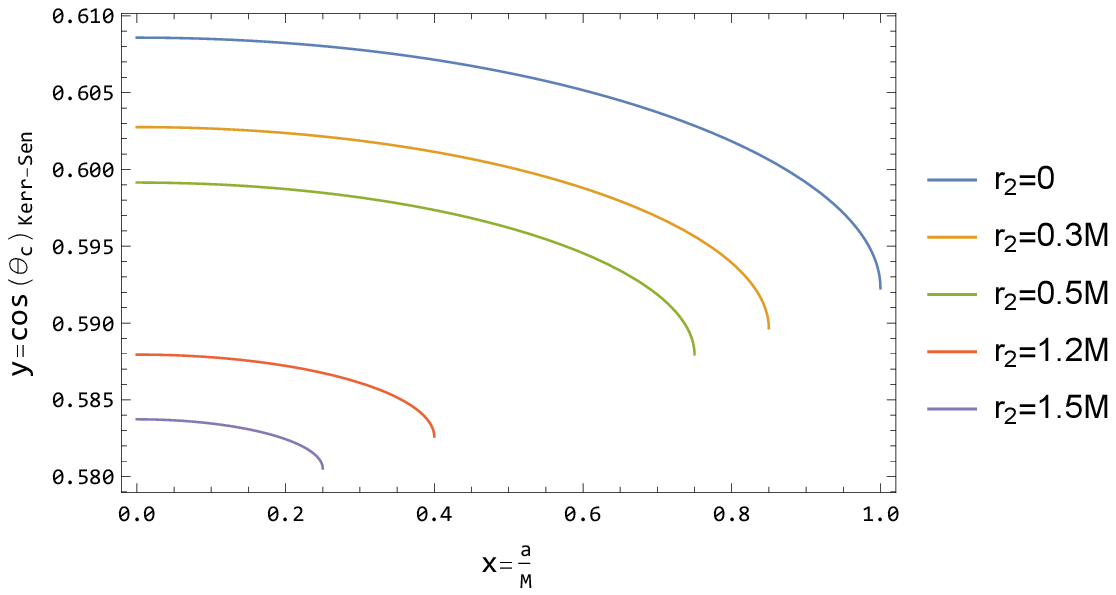}
\end{minipage}
\caption{\label{fig.6}
 For these figures: $V_0=0.4$ and $R=10$ M. The figure in the upper left corner describes the evolution of $\eta$ with respect to $r_2$, and the figure in the upper right corner illustrates the dependence of $a$. The critical angle with variable parameters are displayed in the lower figures.}
\end{figure}

The inflow or outflow of fluid should be accompanied by the injection rate or the ejection rate, respectively. When combined with the mass accretion rate, they constitute the entire concept of choked accretion. These three rates are connected as follows
\be
\label{60}
\dot{\mathscr{M}}_{in}=\dot{\mathscr{M}}+\dot{\mathscr{M}}_{out},
\ee
where the mass accretion rate is decided by Eq. \eqref{32}. It can be described as
\be
\label{61}
\dot{\mathscr{M}}=8\pi M r_{+}\rho_0\Gamma_0\sqrt{\frac{R(R+r_2)\tilde{\Delta}_0}{\tilde{A}_0}},
\ee
For $r_2=0$, Eq. \eqref{61} reverts to the accretion rate of Kerr black hole. It's obvious that the accretion rate is restricted by the parameters chosen and the bounding sphere radius $R$. Now we must identify the critical angle at which the fluid moves towards the black hole in the interval $(\theta_c,\pi-\theta_c)$. In this interval, the projection of the tangent vector of the streamline along the radial direction of the black hole is negative. The critical angle $\theta_c$  could be determined by using  $V^{\hat{r}}$ in \eqref{42}, and the flow is absorbed by the black hole with $V^{\hat{r}}<0$, it follows that
\be
\label{62}
\theta_c=\arccos{\[3\(1-\frac{2Mr_+}{V_0\sqrt{\tilde{A}_0}} \)\]^{-\frac{1}{2}}},
\ee
Then $\dot{\mathscr{M}}_{in}$ associated with the critical angle is
\be
\1\{\begin{split}
\label{63}
&\dot{\mathscr{M}}_{in}=-2\int^{\frac{\pi}{2}}_{\theta_c}\rho U^r\tilde{\Sigma}\sin{\theta}\dif \theta\dif\phi=\frac{8\pi q\rho e M r_+}{h}=q\dot{\mathscr{M}},\\
&q\equiv\frac{2\cos^3{\theta_{c}}}{3\cos^2{\theta_{c}}-1}=\frac{\sqrt{\tilde{A}_0}V_0}{3\sqrt{3}Mr_+}\(1-\frac{2Mr_+}{V_0\sqrt{\tilde{A}_0}} \)^{-\frac{1}{2}}.
\end{split}\2.
\ee
Similarly, the ejection rate could be obtained as $\dot{\mathscr{M}}_{out}=(q-1)\dot{\mathscr{M}}$ by using \eqref{60}. The physics relating to $\dot{\mathscr{M}}_{in}=\dot{\mathscr{M}}$ is that all fluids are dragged into black hole while the critical angle $\theta_c=0$ $(q=1)$. The condition $\theta_c=0$ provides $V_0=\frac{3Mr_+}{\sqrt{\tilde{A}_0}}$ by \eqref{62}. This is in accordance with the constraint from \eqref{51}, as illustrated by the following condition
\be
\label{64}
\dot{\mathscr{M}}_{in}=\left\{
\begin{aligned}
&\dot{\mathscr{M}},\quad V_0\leqslant \frac{3Mr_+}{\sqrt{\tilde{A}_0}},\\
&q\dot{\mathscr{M}},\quad \frac{3Mr_+}{\sqrt{\tilde{A}_0}}\leqslant V_0\leqslant \frac{R(R+r_2)+a^2+6Mr_+}{2\sqrt{\tilde{A}_0}}, \\
\end{aligned}
\right.
\ee
and
\be
\label{65}
\dot{\mathscr{M}}_{out}=\left\{
\begin{aligned}
&0,\quad V_0\leqslant \frac{3Mr_+}{\sqrt{\tilde{A}_0}},\\
&(q-1)\dot{\mathscr{M}},\quad \frac{3Mr_+}{\sqrt{\tilde{A}_0}}\leqslant V_0\leqslant \frac{R(R+r_2)+a^2+6Mr_+}{2\sqrt{\tilde{A}_0}}. \\
\end{aligned}
\right.
\ee
Since the range of $V_0$ changes, $\dot{\mathscr{M}}_{in}$ steadily increases, leading to the maximum value $\theta_{max}$ calculated by the formula below
\be
\label{66}
\frac{1}{3\cos{\theta^2_{max}}}=\frac{R(R+r_2)+a^2+2Mr_+}{R(R+r_2)+a^2+6Mr_+},
\ee
and maximum value of injection accretion rate $\dot{\mathscr{M}}_{inmax}=q_{\max}\dot{\mathscr{M}}$, where
\be
\label{67}
q_{\max}=\frac{2\cos^3{\theta_{max}}}{3\cos^2{\theta_{max}}-1}.
\ee
When $R$ is large, the maximum angle $\theta_{max}\approx54.7^\circ$ ($q_{\max}\rightarrow\infty$) corresponds to the boundary case. To comprehend the transformation of accretion rate ratio, we depicts the ratio of accretion rate $\frac{\dot{\mathscr{M}}_{out}}{\dot{\mathscr{M}}_{in}}\equiv\eta$ from different black holes and $\cos{\theta_c}$ in Figure \ref{fig.6}. In Table \ref{table.2}, we calculate some values for the ratio of injection rate, the ratio of ejection rate, the ratio of accretion rate, $\cos\theta_c$, $\cos\theta_{max}$, and $q_{max}$.

\begin{table}[H]
\renewcommand\arraystretch{1}{
\setlength{\tabcolsep}{2mm}{
\begin{tabular}{|c|cccccccccc|}
\hline
&\multicolumn{10}{|c|}{ $a=0.5$ M, $R=10$ M, $V_0=0.4$} \\
\hline
$\frac{r_2}{M}$ & 0&0.1&0.2&0.3&0.4&0.5&0.6&0.7&0.8&0.9 \\
\hline
$\frac{\dot{\mathscr{M}}_{Kerr-Sen}}{\dot{\mathscr{M}}_{Kerr}}$   &1.000& 0.942& 0.883& 0.824&0.763& 0.702& 0.638& 0.571&0.499&0.418\\
\hline
$\frac{\dot{\mathscr{M}}_{in Kerr-Sen}}{\dot{\mathscr{M}}_{in Kerr}}$    &1.000& 1.007& 1.013& 1.020& 1.026& 1.032& 1.039& 1.045&1.051& 1.057 \\
\hline
$\frac{\dot{\mathscr{M}}_{out Kerr-Sen}}{\dot{\mathscr{M}}_{out Kerr}}$    &1.000& 1.026 & 1.052& 1.078& 1.104& 1.131& 1.160& 1.187&1.217& 1.248\\
\hline
$\cos{\theta_{c}}$    &0.606& 0.604& 0.602& 0.600& 0.598& 0.596& 0.594& 0.592& 0.590& 0.588\\
\hline
$\cos{\theta_{max}}$    &0.598& 0.596& 0.596& 0.594& 0.592& 0.591& 0.590 &0.588& 0.587&0.585\\
\hline
$q_{max}$    &5.950& 6.322& 6.748&7.240& 7.820& 8.513& 9.368& 10.463& 11.956& 14.267\\
\hline
\end{tabular}}}
\caption{\label{table.2}  Assuming that $a=0.5$ M, $R=10$ M, and $V_0=0.4$, some values for the ratio of injection rate, the ejection rate, and the accretion rate, and $\cos\theta_c$, $\cos\theta_{max}$, and $q_{max}$ are shown in the table.       }
\end{table}
In Figure \ref{fig.6}, we illustrate the influence of the parameters $r_2$ and $a$ on the accretion rate ratio for Kerr and Kerr-Sen hole. When $V_0$ and $R$ are fixed, we observe that the ratio $\frac{\eta_{Kerr-Sen}}{\eta_{Kerr}}$ rapidly increases as $r_2$ grows. However, with different fixed values of the dilaton parameter $r_2$, we find that the variation of $\frac{\eta_{Kerr-Sen}}{\eta_{Kerr}}$ with respect to $a$ is not obvious. In Table \ref{table.2}, it also indicates that as the parameter $r_2$ in the Kerr-Sen black hole increases, it amplifies the ejection rate and the injection rate and simultaneously reduces the value of the critical angle and the maximum critical angle. Consequently, this leads to most of the fluid escapes from the black hole, while a very small portion flows into the black hole.

\section{Radiative efficiency and redshift factor from the continuum spectrum}
The continuum spectrum emitted by the accretion disc around a black hole is responsive to the background metric, providing a scheme for observing the Kerr-Sen black hole. In this section, we will calculate the radiative efficiency and the redshift factor from the accretion disc within the permissible interval $0<\frac{r_2}{M}<0.4$. This analysis will allow us to investigate the observable effect of Kerr-Sen black hole and provide a approach to differentiate it from the Kerr black hole. We use the fundamental feature of the Novikov-Thorne model \cite{Novikov:1973kta} provided for delineating the continuum spectrum. The model incorporates several key assumptions: (1) the primary contribution to the continuum spectrum arises from the electromagnetic emission originating from the accretion disk surrounding the black hole; (2) the spacetime around the central massive object is both stationary and axisymmetric; (3) the mass of the accretion disk does not influence the background metric; (4) the accretion disk is characterized as geometrically thin and vertical size is considered negligible compared to its horizontal size; (5) particles around the compact central object traverse between the outer edge ($r_{out}$) and the radius of the innermost stable circular orbit ($r_{isco}$) defined as the inner edge of the disk; (6) the accretion disk is situated in the equatorial plane of the accreting compact object, namely the spin of the black hole perpendicular to the disk surface.

The innermost stable circular orbit radius can be determined from the effective potential $V_{eff}$ \cite{Novikov:1973kta,1974ApJ...191..499P}. For a black hole, the effective potential takes the form
\be
\label{68}
V_{eff}=\frac{E^2 g_{\phi\phi}+2E L g_{t\phi}+L^2 g_{tt}}{g^2_{t\phi}-g_{tt}g_{\phi\phi}},
\ee
where $E$ and $L$ are the specific energy and the specific angular momentum of the test particle and they can be calculated with the following formulas

\be
\1\{\begin{split}
\label{69}
&E=\frac{-g_{tt}-\omega g_{t\phi}}{\sqrt{-g_{tt}-2\omega g_{t\phi}-\omega^2 g_{\phi\phi}}},\\
&L=\frac{g_{t\phi}+\omega g_{\phi\phi}}{\sqrt{-g_{tt}-2\omega g_{t\phi}-\omega^2 g_{\phi\phi}}},\\
\end{split}\2.
\ee
where the angular velocity of the test particle $\omega$ in the equatorial plane ($\theta=\frac{\pi}{2}$) is
\be
\label{70}
\omega=\frac{\dif \phi}{\dif t}=\frac{  -g_{t\phi,r} \pm \sqrt{(-g_{t\phi,r})^2-g_{\phi\phi,r} g_{tt,r} }       }{g_{\phi\phi,r}} .
\ee
The radius of the innermost stable circular orbit $r_{isco}$ corresponds to the inflection point ($V_{eff}=V_{eff,r}=V_{eff,rr}=0$) \cite{Abbas_2023, Heydari-Fard:2022xhr}. Consequently, the determination of $r_{isco}$ from the continuum spectrum could constrain the geometry of spacetime. In particular, if we assume the background to be a Kerr black hole, this measurement can be utilized to predict the angular momentum of the black holes \cite{Brenneman:2013oba}.

\begin{figure}[H]
\centering
\begin{minipage}{0.5\textwidth}
\centering
\includegraphics[scale=0.7,angle=0]{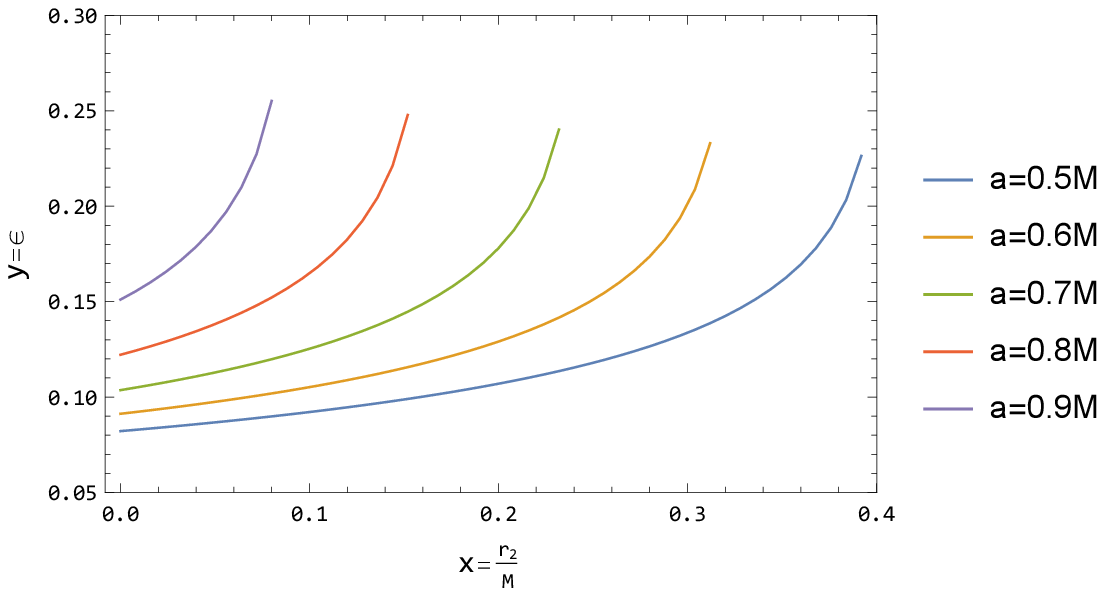}
\end{minipage}%
\begin{minipage}{0.5\textwidth}
\centering
\includegraphics[scale=0.7,angle=0]{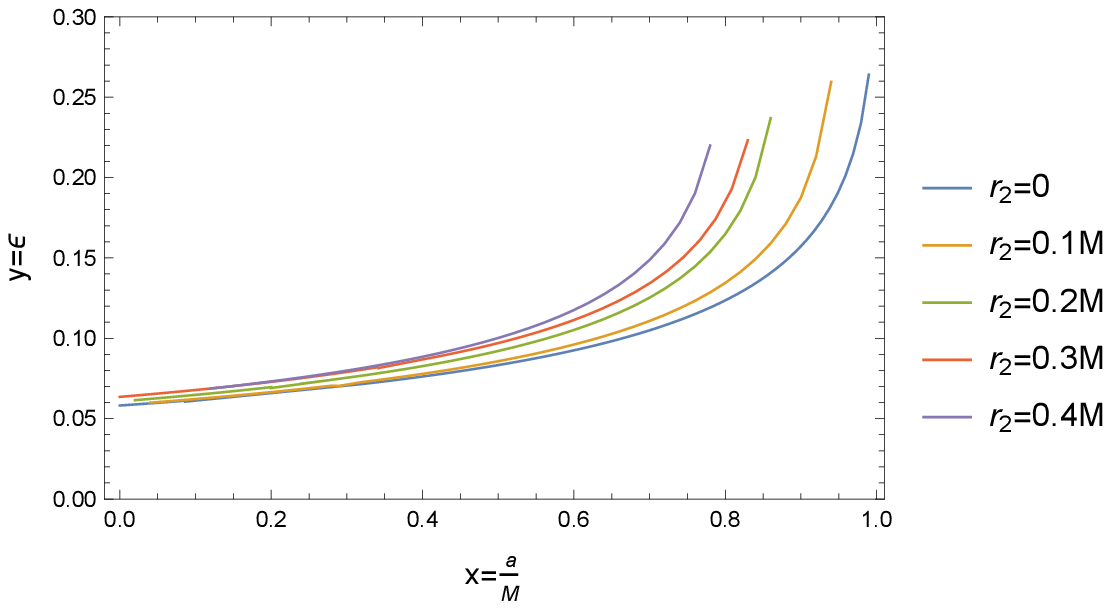}
\end{minipage}
\begin{minipage}{0.5\textwidth}
\centering
\includegraphics[scale=0.7,angle=0]{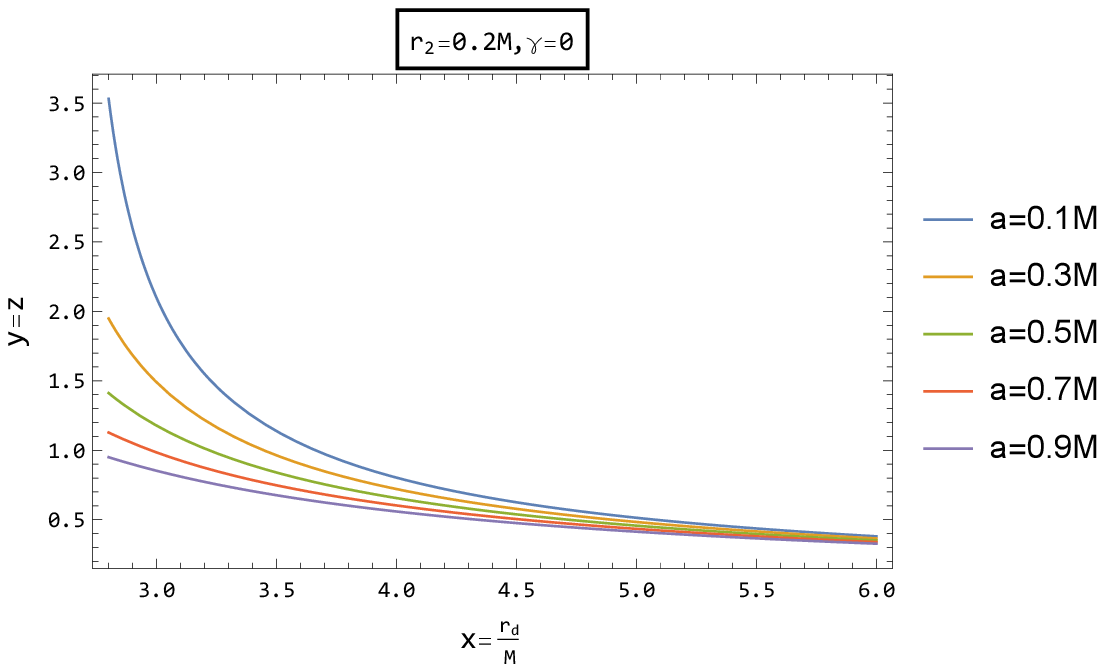}
\end{minipage}%
\begin{minipage}{0.5\textwidth}
\centering
\includegraphics[scale=0.7,angle=0]{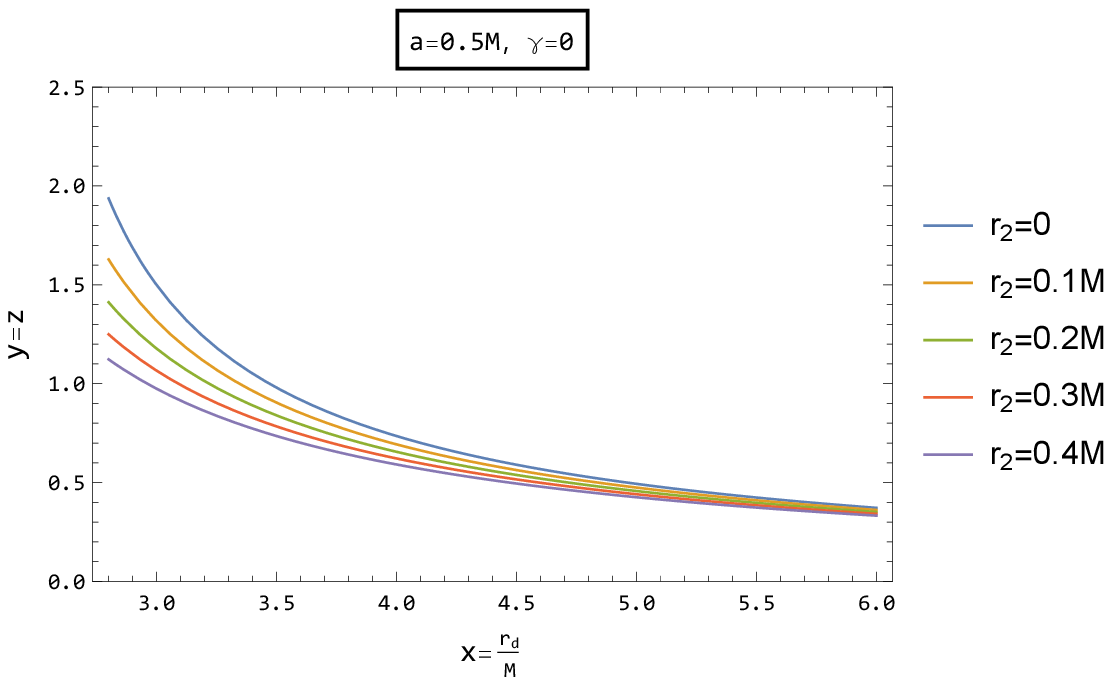}
\end{minipage}

\caption{\label{fig.7}
The upper figures illustrate the variation of the radiative efficiency theoretically derived from the accretion disc. The lower figures illustrate the variation of redshift $z$ with observational distance at $r_2=0.2$ M (or $a=0.5$ M) under an inclination angle of $\gamma=0$.}
\end{figure}

During the accretion process, the efficiency of matter accretion and the redshift factor are important measurement quantities. The redshift factor is associated with the alteration in frequency of a photon as it traverses from the emitting source to the observer. The maximum efficiency $\epsilon$ is determined by the specific binding energy at the marginally stable orbit $r_{isco}$. These two physical quantities are given by
\be
\1\{\begin{split}
\label{71}
&\epsilon=1-E_{isco}(r_{isco}), \\
&g\equiv1+z=\frac{1-\omega r_{d} \sin{\phi}\sin{\gamma}}{\sqrt{-g_{tt}(r_d)-2\omega(r_d) g_{t\phi}(r_d)-\omega^2(r_d) g_{\phi\phi}(r_d)}}, \\
\end{split}\2.
\ee
where $E_{isco}$ is the specific energy of the test particle at $r_{isco}$. The parameters $\gamma$ and $r_d$ represent the inclination angle of the accretion disk and the distance from the observer to the disk, respectively.
The variation of these quantities with parameters are illustrated in Figure \ref{fig.7}. The upper two figures illustrate the variation in radiative efficiency $\epsilon$ with the dilaton parameter $r_2$ (or spin parameter $a$) for different values of $a$ (or $r_2$). The lower figures explain the variations in redshift $z$ with the observational distance with different values of the parameters.

In Figure \ref{fig.7}, we observe that the radiative efficiency of Kerr-Sen black hole increases as the parameter $r_2$ (or the parameter $a$) growing, while the redshift decreases with an increasing observational distance. Furthermore, it is observed that the radiative efficiency of the Kerr-Sen black hole exceeds that of the Kerr black hole with fixed parameter $a$ and the redshift is lower compared to the Kerr black hole. This insight will enhance our understanding of these two black holes through observations.

\section{Conclusion and discussion}
In this work, we initially explored the choked accretion process of ultrarelativistic fluid onto axisymmetric Kerr-Sen black hole in EMDA gravity. Based on the procedure mentioned by Petrich, Shapiro, and Teukolsky, we calculated the solution describing the velocity potential field $\Phi$ of a stationary, irrotational ultrarelativistic fluid in the Boyer-Lindquist coordinates system. We discussed the analytical expression of four-velocity and converted it to the ZAMO framework to give three-velocity. The mass accretion rate, the energy accretion rate, and the angular momentum increase rate were also determined. We found that the increase rate of the angular momentum is zero, indicating that the fluid has an axisymmetric distribution $(m=0)$.

Secondly, as a result of the axial symmetry of perfect fluid and the reflection symmetry in the equatorial plane, we gave the lowest order (2,0) solution, commonly known as quadrupolar flow. We investigated the character of quadrupolar flow and demonstrated that one of the two constraints is automatically satisfied and the timelike requirement of the four-velocity is also resolved once an appropriate region is chosen. Subsequently, we examined the correlation between dilaton parameters $r_2$ and stagnation point $r_s$. From the table, it is evident that increasing dilation parameter leads to a decreasing position of the stationary point of the Kerr-Sen black hole. Then, we introduced the choked accretion model and restricted the physical region as $(r_+, R)$ to ensure that the solution is reasonable. We also calculated the stagnation point analytic formulas based on the boundary values. Within the permissible range of the parameters, the initial velocity $V_0$ at the reference point enables us to depict its dependency on the parameters. The connection between the density ratio at endpoint and at initial point was also evaluated. Additionally, we presented the streamline diagram in cylindrical-like coordinates and discovered that the streamline associated with $\al\leqslant1$ indicates that the fluid was absorbed into the black hole, whereas the streamline with $\al>1$ represents the flow ejected along poles. The injection rate and the ejection rate were discussed in detail at the end of the article. It also indicates that as the parameter $r_2$ in the Kerr-Sen black hole increases, both the ejection rate and the inejection rate are amplified.

Finally, we explored the accretion process in thin disk around the Kerr-Sen black hole by employing the Novikov-Thorne model. Our study focused on analyzing the radiative efficiency and the redshift factor. The results highlight the impact of the dilaton parameter $r_2$ and the spin parameter $a$ on the radiative efficiency and the redshift within the framework of the EMDA model. These results will also contribute to exploring and distinguishing Kerr-Sen black hole from Kerr black hole through observations.

In future work, we will try to solve the choked accretion issue of curled fluid numerically and provide numerical results by merging it with the equation of motion with boundary via the coupling between spacetime and matter field. We will consider the reaction of matter and explore the full accretion issue for dark energy or dark matter as accreted fluid.

\begin{acknowledgments}
This study is supported in part by National Natural Science Foundation of China (Grant No.12333008) and Hebei Provincial Natural Science Foundation of China (Grant No. A2021201034).
\end{acknowledgments}

\appendix
\bibliographystyle{unsrt}
\bibliography{choked._V6}

\end{document}